\documentclass[a4paper,11pt]{article}
\usepackage{jheppub, hyperref, xcolor, slashed, booktabs, diagbox}

\title{Systematic analysis of search strategies for $L_\mu-L_\tau$ gauge bosons at Belle II}

\author[a]{Connor Brown}
\author[a, b]{\!\!, Juri Fiaschi}
\author[a]{\!\!, Oliver Fischer}
\author[a]{\!\!, Thomas Teubner}

\affiliation[a]{Department of Mathematical Sciences, University of Liverpool, Liverpool L69 3BX, United Kingdom}
\affiliation[b]{Universit\`a degli Studi di Milano-Bicocca, Department of Physics “Giuseppe Occhialini”, \& INFN, Sezione di Milano-Bicocca, Piazza della Scienza 3, Milano 20126, Italy}

\emailAdd{connor.brown@liverpool.ac.uk}
\emailAdd{juri.fiaschi@unimib.it}
\emailAdd{oliver.fischer@liverpool.ac.uk}
\emailAdd{thomas.teubner@liverpool.ac.uk}

\abstract{
Extensions of the Standard Model with masses at or below the GeV scale are motivated by searches for dark matter and precision measurements in the quark and lepton flavour sectors, including that of the muon anomalous magnetic moment.
An excellent experimental environment to test such light new physics is given by the Belle II experiment, which foresees to take up to 50~ab$^{-1}$ of data.
Here we consider a model with an additional gauged U(1)$_{L_\mu-L_\tau}$ symmetry that introduces a neutral gauge boson, a Dark Photon, with possibly large couplings to muon- and tau-flavored leptons, including neutrinos.
Dark Photon mixing with the Standard Model photon is loop induced, allowing it to couple to electrically charged fermions other than muons and taus.
We systematically investigate the possible search strategies for Dark Photons with four fermion final states.
We identified search channels with muons as the most promising ones, and we analyse the kinematic distributions to obtain cuts that optimise the sensitivity of Belle II searches for the Dark Photon.
Summarising the sensitivities from the most promising search channels we provide a comprehensive overview of future searches at Belle II.
}

\preprint{LTH-1372}

\keywords{New Light Particles, Specific BSM Phenomenology, Dark Matter at Colliders}

\arxivnumber{2406.10204}

\begin{document} 

\maketitle
\flushbottom

\tableofcontents

\section{Introduction}
The Standard Model (SM) of particle physics is a remarkably successful theory, thoroughly tested by many experiments in the past decades.
The theory appears incomplete as it is unable to explain numerous phenomena - such as the existence of dark matter, the baryon asymmetry of the universe, and neutrino masses - while the increasing precision of modern measurements has revealed significant tensions - as the recent measurements 
of $g-2$, the muon anomalous magnetic moment, at Fermilab~\cite{Muong-2:2021ojo, Muong-2:2023cdq, Muong-2:2024hpx}, or the measurement of the $W$-boson mass by the CDF collaboration at the Tevatron~\cite{CDF:2022hxs}.
New physics below the electro-weak (EW) scale, coupling very weakly to the SM, could solve tensions between SM predictions and experimental observations, including anomalies in the flavour sector.
Importantly, new light states, which are singlet under the SM gauge group, can play the role of portals to hidden or dark sectors, hence connecting ordinary and dark matter.

A minimal extension of the SM, connecting the visible to a dark sector, can be constructed by adding an U(1)$_X$ Abelian gauge group which mixes with the SM U(1)$_Y$ hypercharge field~\cite{Bauer:2022nwt}.
The $L_\mu - L_\tau$ gauge symmetry is a popular choice for the new Abelian group, as it naturally predicts a small kinetic mixing with the SM neutral gauge bosons, while keeping the theory renormalisable and anomaly free~\cite{Essig:2009nc}.
Such lepton non-universal interactions could also - at least partially - account for the discrepancy between theory~\cite{Aoyama:2020ynm,Colangelo:2022jxc} and measurements~\cite{Muong-2:2021ojo, Muong-2:2023cdq} of the muon anomaly. 
Currently, the status of the $g-2$ discrepancy is unclear, due to significant tensions between data used in~\cite{Aoyama:2020ynm} as input in the data-driven evaluation of the hadronic vacuum polarisation contributions to $g-2$ and the recent two-pion data from CMD-3~\cite{CMD-3:2023alj,CMD-3:2023rfe}, and due to tensions with the lattice predictions, in particular from the BMW collaboration~\cite{Borsanyi:2020mff}. However, with a lot of efforts on new and improved low-energy hadronic cross section measurements and on further lattice determinations, the comparison between the $g-2$ measurement and its SM prediction is expected to be consolidated and could still indicate a significant discrepancy, or, in turn, strongly constrain physics beyond the SM.

Neutral vector bosons with masses below the GeV scale, generally called Dark Photons (DPs), can be probed looking at final states involving photons and leptons~\cite{Curtin:2014cca}.
Experiments at colliders provide the strongest constraints for DP masses in the MeV to GeV range~\cite{Ilten:2018crw}.
The LHC main experiments have performed dedicated DP searches, see e.g.\ LHCb~\cite{LHCb:2017trq, LHCb:2019vmc}, ATLAS~\cite{ATLAS:2019tkk, ATLAS:2022xlo}, CMS~\cite{CMS:2023hwl, CMS:2020krr, CMS:2019ajt, CMS:2018yxg}. However, due to the leptophilic nature of the $L_\mu - L_\tau$ symmetry, lepton colliders present unique opportunities to test this model.
Because of the vanishing DP tree-level couplings to coloured particles, purely leptonic final states with high multiplicity are the most promising channels for investigating this model.
In order to probe very low couplings, high intensity beams are needed to produce DPs at a detectable rate.
High luminosity experiments like BaBar~\cite{BaBar:2014zli, BaBar:2016sci}, KLOE~\cite{KLOE-2:2014qxg}, Belle~\cite{Belle:2021feg} and Belle II~\cite{Ban:2020uii, Belle-II:2023ydz} at the meson factories DA$\Phi$NE, PEP-II and KEK, are able to collect very large data samples, hence providing the highest sensitivity for $L_\mu - L_\tau$ DPs.

The DP associated with the U(1)$_{L_\mu - L_\tau}$ symmetry group can also act as a portal to a dark sector through its direct couplings to new invisible light states, such as dark chiral fermions or sterile/Majorana right-handed neutrinos, which can (at least partially) account for the observed dark matter relic abundance.
DP portals to dark sectors have been probed through signatures with missing energy at lepton colliders (BaBar~\cite{BaBar:2017tiz}) and beam dump experiments (E137~\cite{Batell:2014mga}, NA64~\cite{NA64:2024klw}), or through their interactions with neutrinos (in indirect detection at IceCube~\cite{Kamada:2015era, Hooper:2023fqn}, in neutrino trident production~\cite{Altmannshofer:2014pba}, at LSND~\cite{deNiverville:2011it}, CHARM-II~\cite{CHARM-II:1990dvf}, CCFR~\cite{CCFR:1991lpl}, Borexino~\cite{Gninenko:2020xys, Bellini:2011rx}, see also~\cite{Amaral:2020tga}) and dark matter (XENON~\cite{Essig:2012yx, Essig:2017kqs}) in scintillator and direct detection experiments~\cite{Figueroa:2024tmn}.
Dark matter coupled to the DP carries complementary constraints~\cite{Aboubrahim:2022qln} from astrophysics and cosmology~\cite{Foldenauer:2024cdp, Giovanetti:2021izc, Escudero:2019gzq, Foldenauer:2019dai, Kamada:2018zxi, Bauer:2018onh, Feng:2016ijc} as well as indirect detection (FERMI-LAT~\cite{Fermi-LAT:2015att}) surveys.

Light DPs weakly coupled to the SM naturally feature macroscopic decay lengths, which lead to detectable displaced decay vertices.
This is the case if the DP has a narrow width, which still holds in models where the DP acts as a portal to an extended dark sector, as long as the dark particles are heavy enough and any additional decay channels for the DP are kinematically forbidden.
Signatures with displaced vertices represent a smoking gun for the discovery of long-lived particles.
A very active ongoing programme at the LHC focuses on improving the sensitivity on long-lived particles~\cite{LHCb:2020ysn, Lee:2018pag, Alimena:2019zri, Bernreuther:2020xus}, also exploiting the CERN Super Proton Synchrotron beam dump (NA64++~\cite{Gninenko:2014pea, Gninenko:2018tlp, Gninenko:2018num}, NA48/2~\cite{NA482:2015wmo}), the muon beam (MUonE~\cite{Galon:2022xcl}), or novel dedicated forward detectors (FASER~\cite{Feng:2017uoz, Feng:2018pew, FASER:2018eoc}, MATHUSLA~\cite{Curtin:2017izq, Curtin:2018mvb}, CODEX-b~\cite{Gligorov:2017nwh}).

Similarly, dedicated searches for displaced signatures at lepton colliders~\cite{Bossi:2013lxa} have been pursued by the KLOE~\cite{KLOE-2:2018kqf}, BaBar~\cite{BaBar:2015jvu, BaBar:2016sci}, and Belle II~\cite{Belle-II:2010dht, Duerr:2019dmv, Duerr:2020muu, Bernreuther:2022jlj, Ferber:2022ewf, Belle-II:2023ueh} experiments. 
In the future, the light long-lived DP scenario can be best tested in the clean environment of $e^+ e^-$ collisions with the Belle II detector.
Thanks to its large acceptance and with the prospect of high luminosity in the forthcoming years, we can expect a statistically significant sample to study long-lived DP signatures.
The high energy asymmetric $e^+ e^-$ collisions generate a boost of the centre-of-mass frame, enhancing the lifetime of the final state system and the probability to observe events with a displaced decay vertex.
The recast of the experimental displaced vertices search in the DP framework therefore provides a powerful probe to rule out yet unexplored regions in the parameter space for such BSM scenarios.

In this paper we consider the sensitivity of the Belle II target luminosity on the minimal BSM construction featuring a DP from an additional U(1)$_{L_{\mu}-L_{\tau}}$ gauge group, connected to the SM through kinetic mixing generated at loop-level (for studies on  less constrained model see e.g.~\cite{Bandyopadhyay:2022klg, Bandyopadhyay:2023lvo}).
In Sect.~\ref{sec:model} we introduce the DP model and its phenomenological key aspects, and discuss existing constraints.
In Sect.~\ref{sec:signatures} we assess the potential of DP searches at Belle II, with detailed analyses of the relevant signatures involving the new light neutral boson, including displaced vertex searches for long-lived DPs.
In Sect.~\ref{sec:analysis_methodology} we describe our computational framework and statistical method to derive experimental sensitivity bounds.
In Sect.~\ref{sec:searches} we systematically analyse background and signal rates in the relevant final states.
In Sect.~\ref{sec:results} we project the sensitivity of the ultimate Belle II target luminosity on the parameter space of the model, and compare it with the existing constraints.
Finally, in Sect.~\ref{sec:conclusions} we draw our conclusions.

\section{Dark Photon Model and Decays}
\label{sec:model}
We introduce a minimal BSM construction extending the SM gauge group with the local abelian group U(1)$_{L_\mu - L_\tau}$, with $L_\alpha$ being the lepton number of flavour $\alpha$~\cite{Heeck:2011wj}. 
The corresponding gauge boson $\hat X^{ \mu}$ interacts with the SM lepton current $j_\mu^{L_\mu - L_\tau}$, i.e.\ with the SM muon and tau charged and neutral leptons.
The effective Lagrangian after EW symmetry breaking contains the terms
\begin{equation}
    {\cal L} \ \supset\ -\frac{1}{4}  \hat F_{\mu\nu} \hat F^{\mu\nu} -\frac{1}{4} \hat X_{\mu\nu} \hat X^{ \mu\nu}  -\frac{\epsilon}{2} \hat F_{\mu\nu} \hat X^{\mu\nu} - \frac{1}{2} m_X \hat X^\mu \hat X_\mu\,,
    \label{eq:lagrangian}
\end{equation}
where $\hat F_{\mu\nu}$ and $\hat X_{\mu\nu}$ are the field strength tensors corresponding to the SM photon field $\hat A_\mu$ and the dark photon $\hat X_\mu$, respectively.
\begin{figure}
    \centering
    \includegraphics[width=0.60\textwidth]{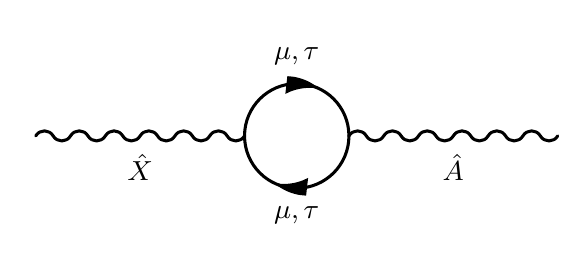}
    {\vspace{-0.6cm}}
    \caption{Loop induced mixing of the dark photon $\hat{X}$ with the SM photon $\hat{A}$.}
    \label{fig:loop-mixing}
\end{figure}
The U(1) kinetic mixing term in the Lagrangian effectively arises, at loop level, from the interaction between the $\hat X$ boson and the $j_\mu^{L_\mu - L_\tau}$ current, as illustrated in Fig.~\ref{fig:loop-mixing}. While in the general case the kinetic mixing parameter $\epsilon$ can be treated as a free parameter of the Lagrangian, in this specific construction it is given by \cite{Bauer:2018onh}
\begin{equation}
\epsilon = \frac{e g_X}{2\pi^2} |f(m_\mu,m_\tau,q)|\,,\label{eq:loop-mixing}
\end{equation}
with the loop factor 
\begin{equation}
    f(m_\mu,m_\tau,q) \ =\ \int\limits_0^1 dx\,  x (1 - x)\ln\left.\frac{m_\mu^2 - q^2 x(1-x)}{m_\tau^2 - q^2 x(1-x)}\right.\,. 
\label{eq:loop-factor}
\end{equation}
The explicit mass $m_X$ of the DP may arise e.g.\ from higher scale interactions with an extended Higgs sector.
In the following we will assume $m_X$ in the few MeV to few GeV range.
We do not consider additional explicit mass mixing with the much heavier $Z$-boson field, which is strongly constrained by electroweak precision measurements~\cite{Heeck:2011wj, Babu:1997st}.
Non-diagonal elements in the mass matrix inherited from the small kinetic mixing $\epsilon$ can also be safely neglected.
With this choice, the mass eigenstates ($\hat A, \hat X$) are diagonal in the interaction basis ($A, X$).

Theoretical consistency requires anomaly cancellation. Even after adding right-handed neutrinos fields, which allow for the observed neutrino oscillations, this minimal construction is automatically anomaly free.
We give Dirac masses of ${\cal O}(1)$~eV to muon and tau neutrinos, but do not introduce Majorana masses, such that the mass eigenstates remain light.
While with this choice the following results remain general, careful fine-tuning of the parameters of the neutrino sector - possibly also requiring right-handed neutrinos in the spectrum - is needed to satisfy the experimental constraints (see Ref.~\cite{Asai:2017ryy, Bauer:2018onh}).

\begin{figure}
    \centering
    \includegraphics[width=0.48\textwidth]{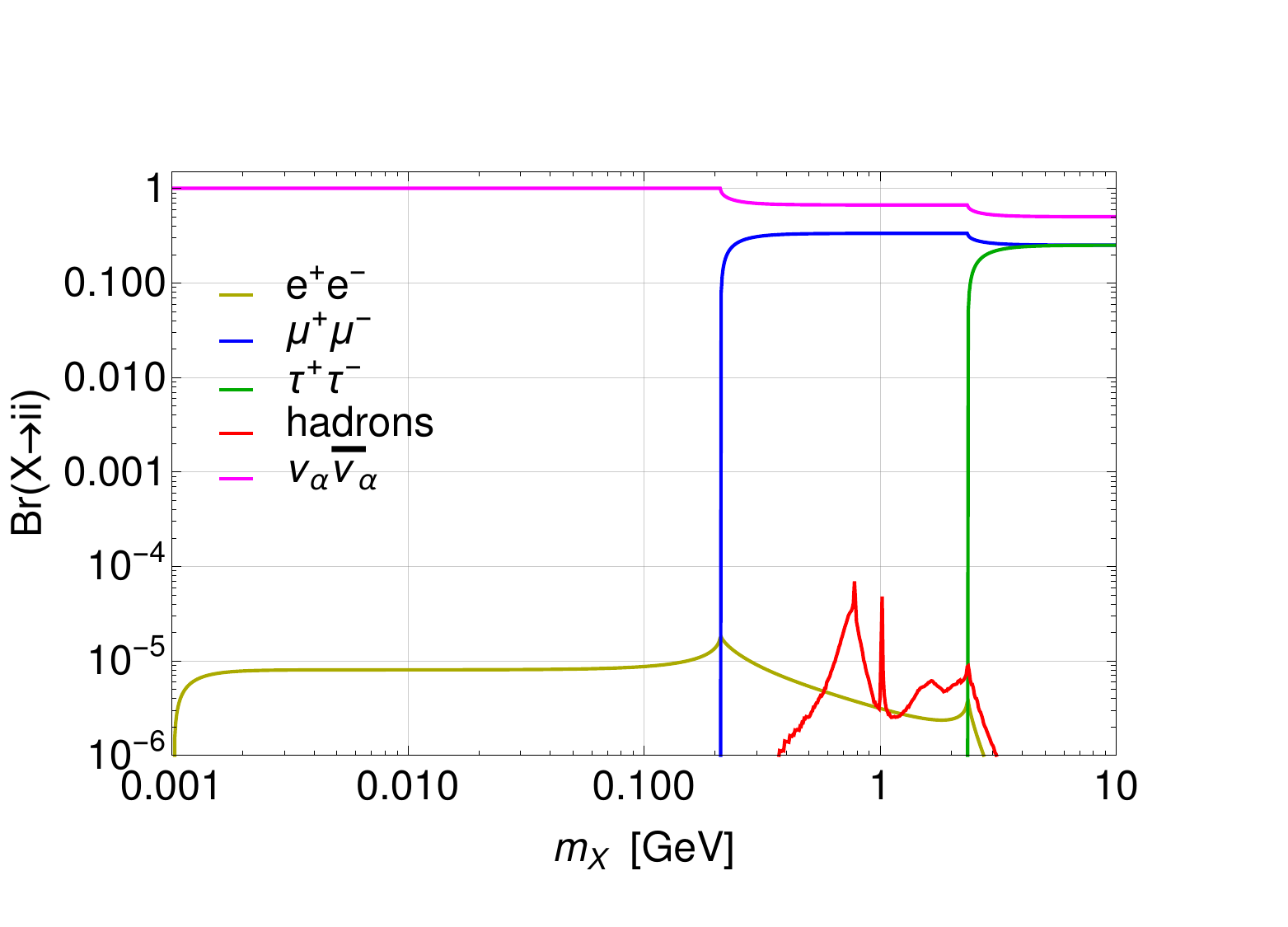}
    \includegraphics[width=0.48\textwidth]{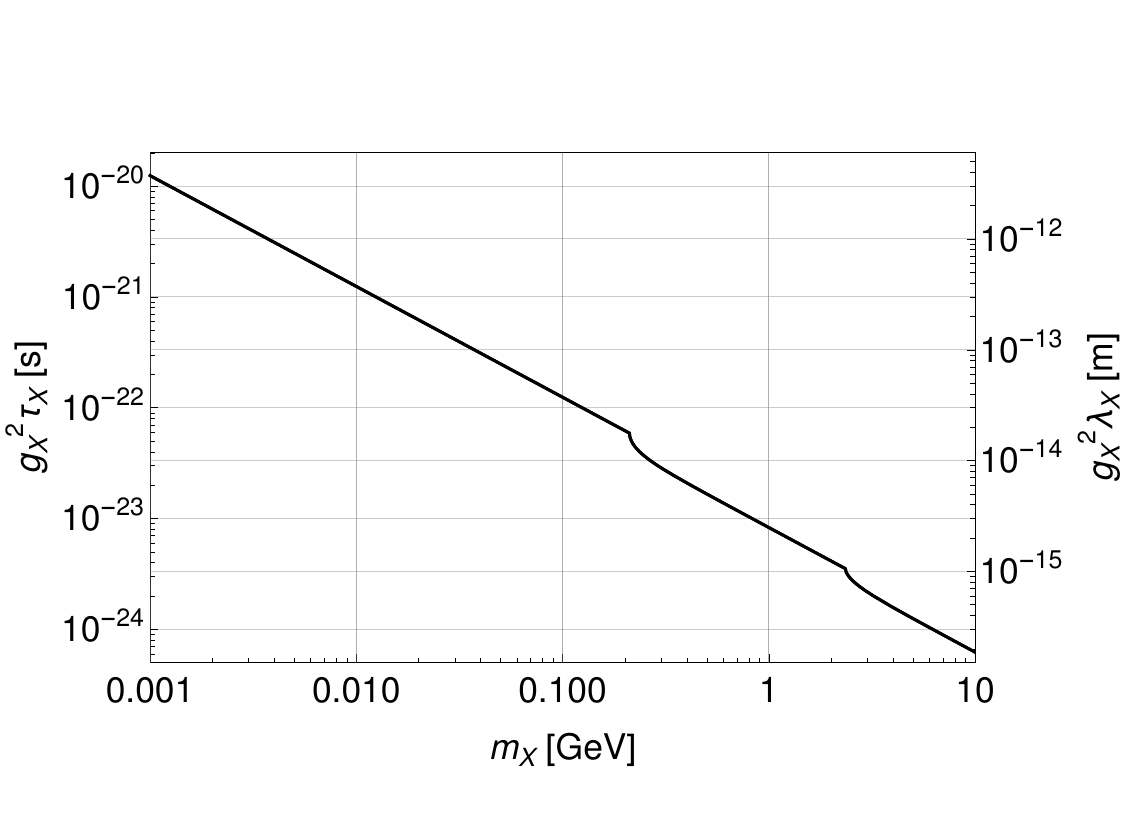}
    {\vspace{-0.6cm}}
    \caption{Branching ratio of DP as a function of its mass (left), and DP mean lifetime and decay length in the DP rest frame multiplied by $g_X^2$ as a function of its mass (right).}
    \label{fig:Branching_ratio}
\end{figure}

The DP couples directly to muon and tau lepton flavours with coupling strength $\alpha_X = g_X^2 / (4\pi)$, and to all electrically charged particles with effective coupling strength $\epsilon^2\alpha_{\rm em}$. It decays into all fermion anti-fermion pairs with fermion masses below $m_X/2$, with a partial decay width
\begin{equation}
    \Gamma_i \ =\ \frac{1}{3} m_X (\delta_{i,(\mu,\tau)}\alpha_{\rm X} + \epsilon^2 \alpha_{\rm em} q_i^2 N_c) \left(1 + \frac{2 m_i^2}{m_X^2}\right)\sqrt{1 - \frac{4 m_i^2}{m_X^2}} \,\theta(m_X - 2 m_i)\,,
\end{equation}
where $m_i$ and $q_i$ denote the fermion's mass and charge respectively, and $N_c$ is the number of colours (1 for leptons and 3 for quarks).
The $\delta_{i,(\mu,\tau)}$ indicates that the tree-level contribution is present only for decays into muon and tau leptons.
The interference between the $L_\mu - L_\tau$ and electromagnetic currents from kinetic mixing is small and will be neglected.
The DP decay into neutrino states (both left- and right-handed), neglecting their mass, also occurs at tree-level via its coupling to $\nu_\mu$ and $\nu_\tau$ flavours:
\begin{equation}
    \Gamma_{\nu_i} \ =\ \frac{1}{3} m_X \delta_{\nu_i,(\nu_\mu,\nu_\tau)} \alpha_{\rm X}\,.
\end{equation}
The left panel of Fig.~\ref{fig:Branching_ratio} shows the DP branching ratios as a function of its mass, clearly demonstrating the strong suppression of decay channels that proceed via kinetic mixing compared to those into fermions charged under $L_\mu-L_\tau$. 
The DP hadronic branching ratio is obtained by appropriately rescaling the experimental data for the ratio $R = \frac{\sigma (e^+e^- \rightarrow \rm{hadr.})} {\sigma (e^+e^- \rightarrow \mu^+\mu^-)}$~\cite{ParticleDataGroup:2022pth}.
Its contribution is small and irrelevant for our analysis, as the DP width is dominated by decays into muon and tau flavoured leptons above their mass threshold, and by decays into electrons below 500~MeV.

The mean lifetime $\tau_X$ and decay length $\lambda_X$ of the DP in its rest frame are calculated as:
\begin{equation}
    \tau_X = \frac{\lambda_X}{c} = \frac{\hbar}{\Gamma_X},\qquad \text{with}\qquad \Gamma_X = \sum\limits_s \Gamma_i
    \label{eq:lifetime}
\end{equation}
and are shown in the right panel of Fig.~\ref{fig:Branching_ratio}; in the plot we show the quantities $g_X^2 \tau_X$ and $g_X^2 \lambda_X$ which are independent of $g_X$.

\section{Dark photon signatures at Belle II}
\label{sec:signatures}
Dark photons with masses between MeV and GeV are testable in low energy $e^+e^-$ collisions.
In the following, we focus on associated DP production at the Belle II experiment via the processes $e^+e^- \to X Y$, where $Y=\gamma, f^+f^-$ and $f=e,\mu,\tau$ or hadrons.
Belle II is a general purpose detector at the $e^- e^+$ collider KEKB which runs at the centre-of-mass energy $\sqrt{s}$ = 10.58~GeV with an anti-symmetric beam energy setup (E($e^-$) = 7~GeV, E($e^+$) = 4~GeV). Operating mostly at the $Y(4S)$ resonance, its primary focus is to study $B$ meson properties~\cite{Belle-II:2018jsg}.
The Belle II detector covers an asymmetric phase space, with the polar angle spanning from $\theta_l = 17^\circ$ to $\theta_u = 150^\circ$, with the positive $z$-axis in the direction of the $e^-$ beam.
The Central Drift Chamber (CDC) has an inner tracker radius $R_T$ = 0.168~m (with $R_T^2 = R_x^2 + R_y^2$) and barrel length $R_z$ = 1.15~m.
Its current data set corresponds to an integrated luminosity of $L$ = 427~fb$^{-1}$, while the ultimate target luminosity is $L$ = 50~ab$^{-1}$~\cite{Bertacchi:2023jzv}.

\begin{figure}
    \centering
    \includegraphics[width=0.90\textwidth]{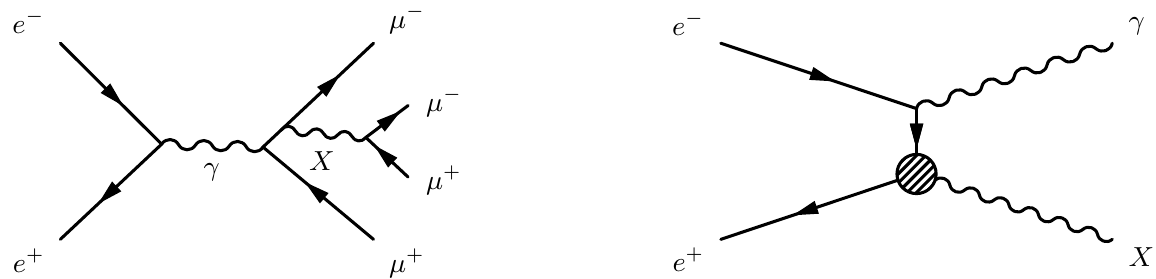}
    \caption{Feynman diagrams depicting the considered DP production processes at $e^+e^-$ colliders. \textit{Left:} Radiation off a muon or tau lepton. \textit{Right:} Mono photon process.}
    \label{fig:feynman}
\end{figure}

\begin{figure}
    \centering
    \includegraphics[width=0.7\textwidth]{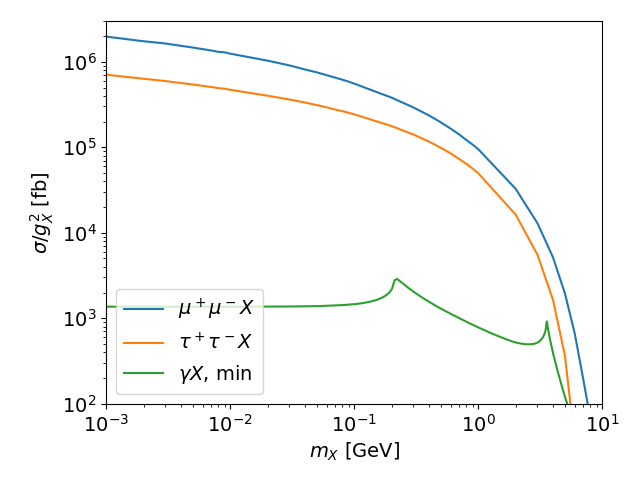}
    {\vspace{-0.4cm}}
    \caption{Dark photon production cross sections at Belle II (with $\sqrt{s}=10.58$~GeV), divided by the U(1)$_{L_\mu-L_\tau}$ coupling constant, $g_X^2$, for the production mechanisms in $e^+e^-$ collisions as described in the text. 
    The Belle II geometric detector acceptance has been imposed on the leptons and the photon, as well as a minimum energy cut of 1~GeV for the photon.}
    \label{fig:darkphoton_production_belle}
\end{figure}

In the following we discuss the relevant experimental signatures in the Belle II environment which are sensitive to DPs within the model introduced in Sect.~\ref{sec:model}.
The final states considered comprise combinations of photons, leptons with prompt or displaced decay vertices, and of missing transverse energy/momentum originating from invisible DPs decays.
Because of the DP's suppressed coupling to electrons, cf.\ Eq.~\eqref{eq:loop-mixing}, and the large SM background, we do not consider resonant processes with two body final states, $e^+e^-\to X\to f^+f^-$. 
Nevertheless, DPs can be produced through their unsuppressed couplings to muon and tau leptons, via the processes $e^+e^- \to \ell^+\ell^- X$ with $\ell = \mu,\tau$, i.e.\ when radiated off a heavy lepton final state.
They can also be produced through kinetic mixing in association with a SM photon, via the one-loop induced process $e^+e^- \to \gamma X$, with amplitude proportional to $\epsilon^2$.
The corresponding Feynman diagrams are shown in Fig.~\ref{fig:feynman}.
Because of their very narrow width, we will exclusively consider DP on-shell production and set the momentum transfer to $q^2 = m_X^2$ when evaluating $\epsilon$ in Eq.~\eqref{eq:loop-mixing}.
For example, in the regime where $m_X \ll m_\mu$, 
we obtain $\epsilon \simeq 0.015\,g_X$, whereas $m_X = 10$~GeV yields $\epsilon \simeq 6\times10^{-4}g_X$.
In Fig.~\ref{fig:darkphoton_production_belle} the cross sections, divided by the coupling constant, $g_X^2$, for the three main production mechanisms, are shown as function of the DP mass in the Belle II detector acceptance region, including also a lower cut on the photon energy of 1~GeV.

The relevant processes for DP searches at Belle II are obtained combining the decay channels of the DP $X$, depicted in the left panel of Fig.~\ref{fig:Branching_ratio}, with the possible associated particles $Y$.
Among the possible combinations, processes where both $Y$ and the DP decay involve a combination of $\mu,\tau,\nu$ fermions are most relevant, due to their large production cross section and ${\cal O}(1)$ branching ratio.
These signatures are expected to give the largest sensitivity to DP searches and could become ``discovery channels''.
In this group, we also include the so-called ``monophoton'' channel, where $Y = \gamma$ and the DP decays invisibly, see e.g.~\cite{Kaneta:2016uyt, Araki:2017wyg} for earlier works. This signature is somehow exceptional because, as we will discuss below, excellent background suppression can be achieved, and background free analyses are justified.
For completeness we also consider the associated production of a DP with $Y=\nu\bar\nu$.
This process occurs via weak interactions and is suppressed by a factor $r_W = s^2 G_f^2 \simeq 10^{-6}$ compared to QED processes with Fermi's constant, $G_f = 1.166 \times 10^{-5}$~GeV$^2$~\cite{ParticleDataGroup:2022pth}.
The cross sections for processes where $Y \neq \mu,\tau,\nu$ are loop-suppressed by a factor $r_\epsilon = \epsilon^2/g_X^2 < 10^{-4}$ compared to the discovery channels, and are therefore expected to provide lesser sensitivity.
Final states stemming from DP decays into electrons or hadrons have branching ratios loop-suppressed by the same factor $r_\epsilon$, therefore their cross sections are much smaller compared to the discovery channels.
The discussed final states are summarised in Tab.~\ref{tab:signal_channels}, with their corresponding suppression factors with respect to the SM QED background.
\begin{table}
    \centering
\begin{tabular}{c|ccc}
    \backslashbox{$Y$}{$X\to $} & $\nu$ & $e$/hadrons & $\mu/\tau$  \\
    \hline
    $\nu$ & invisible & $r_W r_\epsilon$ & $r_W$ \\  
    $e/\gamma/$hadrons & $r_\epsilon$ & $r_\epsilon^2$ & $r_\epsilon$ \\
    $\mu/\tau$ & 1 & $r_\epsilon$ & 1 \\
\end{tabular}    
    \caption{Four fermion final states from the process $e^+e^-\to XY$ as discussed in the text. Entries denote suppression factors due to the loop-induced coupling $\epsilon$, or weak interactions. Leading processes are labelled ``1''.}
    \label{tab:signal_channels}
\end{table}

Existing Belle II searches for light DPs mainly target final states with missing energy from the invisible decays of the DP.
Analyses of the process $e^+e^- \to \mu^+\mu^-X$ with $X\to$ invisibles have been performed using a dataset corresponding to an integrated luminosity of 276~pb$^{-1}$~\cite{Belle-II:2019qfb} and, more recently, of 79.7~fb$^{-1}$~\cite{Belle-II:2022yaw}, achieving exclusion limits between $g_X \simeq 3\times10^{-3}$ for DPs with $m_X \ll 1$~GeV and $g_X \simeq 1$ for DPs as heavy as 8~GeV.
For the ultimate Belle II integrated luminosity goal of 50~ab$^{-1}$, an exclusion limit on the DP coupling down to $g_X \simeq {\cal O}(10^{-4})$~\cite{Belle-II:2019qfb} is projected for this channel.
Analyses targeting the specific $L_\mu - L_\tau$ gauge symmetry, exploiting alternative signatures, such as the monophoton channel with an assumed integrated luminosity of 20~fb$^{-1}$~\cite{DePietro:2018sgg}, and final states with DPs decaying into muon or tau pairs with an assumed integrated luminosity of 100~fb$^{-1}$~\cite{Laurenza:2022rjm}, have demonstrated a promising sensitivity, comparable with the $2\mu + \slashed{E}$ final state.

Previous studies from the BaBar collaboration also reported searches for new light physics in final states with photons and leptons.
They probed a general model with kinetic mixing between the SM photon and DPs, using the processes $e^+e^- \to \gamma \ell^+\ell^-$ with $\ell = e,\,\mu$, based on their complete dataset corresponding to a total luminosity of 514~fb$^{-1}$.
Upper limits at 90\% confidence level (CL) on $\epsilon$ were drawn between $10^{-4}$ and $10^{-3}$ for DP masses between 0.02~GeV and 10.2~GeV~\cite{BaBar:2014zli}.
In the same mass range, a dedicated analysis on SM extensions with extra U(1)$^\prime$ $L_\mu - L_\tau$ gauge symmetry, using the four muon final state~\cite{Godang:2016gna}, yielded the current, most stringent upper bound, $g_X = 7\times 10^{-4}$.

\section{Analysis setup}
\label{sec:analysis_methodology}
We have implemented the minimal DP model introduced in Sect.~\ref{sec:model} into \texttt{Feynrules}~\cite{Christensen:2008py} and generated the model file for the Monte Carlo event generator \texttt{WHIZARD}~\cite{Kilian:2007gr, Moretti:2001zz, Christensen:2010wz}.
We then used \texttt{WHIZARD} to generate samples containing between $10^5$ and $10^7$ events, and to calculate cross sections for leading order signal and background processes.
The complete information on the kinematics of each event is recorded in the four-vectors of the final state particles which are then used to compute the relevant observables for each of the analyses described in the following.

The finite resolution of the Belle II detector for particles' momenta and energies is modelled by smearing the four-momenta following a normal distribution with standard deviations of 0.5\% for muons and electrons, and of 5\% for photon energies~\cite{Ferber:2022rsf}.
The energy resolution for charged particles is taken to be 5~MeV.
Particle identification is highly important in order to study different event topologies at Belle II~\cite{Kuhr:2018lps}.
We consider the reconstruction efficiencies of photons as 100\% for $E_A = {\cal O}(1)$~GeV, of muons $\sim 90\%$ and of electrons $\sim 95\%$~\cite{Belle:2021feg, Belle-II:2021rni}.
Contributions from mis-reconstructed particles are neglected, since for muons they are as low as 0.9\%, while the pion fake rate is 1.4\%~\cite{Abashian:2002bd}.
Complicated backgrounds such as decay chains of $D$ mesons can also lead to the mis-identification of pions as muons or electrons, with a likelihood as high as 10\% for pion momenta $P_\pi \leq 3$~GeV.
This contribution is not included in our analyses, since we do not include QCD processes which are suppressed in this model.

The SM QED processes representing the background for DP searches and their fiducial cross sections are listed in Tab.~\ref{tab:SM_background_processes}. 
\begin{table}
    \centering
    \begin{tabular}{c|c|c|c}
           signature & process & final state & $\sigma_b$ \\
            \hline
         & di-photon & $\gamma\gamma$ & 0.167 nb \\ 
        monophoton & radiative Bhabha scattering & $\gamma e^+e^-$ & 6.28 nb\\
        & radiative neutrinos & $\gamma\bar\nu\nu$ & 2.41 fb\\
            \hline
        & di-tau & $\tau^+\tau^-$ & 0.802 nb \\
        $2\mu + \slashed{E}$ & radiative di-muon & $\gamma\mu^+\mu^-$ & 0.323 nb\\
        & di-electron di-muon & $e^+e^-\mu^+\mu^-$ & 14.9 nb \\
            \hline
        $4\mu$ & four muon & $2\mu^+2\mu^-$ & 155 fb \\
        & di-tau di-muon & $\tau^+\tau^-\mu^+\mu^-$ & 130 fb \\
    \end{tabular}
    \caption{SM QED background processes for DP searches, grouped by the associated signature. The cross sections are computed applying the cuts described in Sect.~\ref{sec:searches}, and include a minimum energy cut of 1~GeV for the photons.}
    \label{tab:SM_background_processes}
\end{table}
Most search channels are affected by enormous SM background rates which dominate the signatures for DP signals.
In the following analyses we find optimised kinematic cuts to enhance the signal-over-background ratio. Specific observables are considered to discriminate the DP signals from the background. These are used to extract the experimental sensitivity on the model parameters at different luminosity stages.
When background rates are much larger than the signal, $N_b \gg N_s$, the optimised combination of kinematic cuts for each phase space region is identified by minimising the critical number of signal events for which the BSM hypothesis is excluded at a certain Confidence Level (CL):
\begin{equation}
N_s^{\rm crit} = \frac{n\,\left(\sum_i \sigma_i \epsilon_i \, L\right)^{\frac{1}{2}}}{\epsilon_s}\,,
    \label{eq:limit-gauss}
\end{equation}
where the sum is over all contributing background processes with their respective cross sections $\sigma_i$ and selection cut efficiencies $\epsilon_i$. 
We choose $n=2$ for a 95\% one-sided CL, and $\epsilon_s$ is the overall selection cut efficiency for the signal.
When the number of background events is less than ${\cal O}(100)$, we evaluate the critical number of signal events assuming a Poisson distribution, and calculate $N_s^{\rm crit}$ as
\begin{equation}
    \sum\limits_{k=0}^{N_b} \frac{\lambda^k e^{-\lambda}}{k!}\ =\ 1-CL\,,
    \label{eq:limit-poisson}
\end{equation}
with $\lambda = N_s^{\rm crit} + N_b$, and CL = 95\%.
In the special case of a completely reducible background, i.e.\ in background free analyses, we obtain $N_s^{\rm crit} = 2.99$.
From $N_s^{\rm crit}$ we extract the critical signal cross section by $\sigma_s^{\rm crit} = N_s^{\rm crit}/(\epsilon_s L)$, from which in turn we obtain the critical value of $g_X$ above which the signal is observable at the chosen confidence level.

\section{Dark photon search channels}
\label{sec:searches}
In the following we describe optimised analyses for each relevant search channel for DP detection at Belle II, i.e.\ for monophoton, two muon plus missing energy, four muon prompt, and four muon displaced signatures. In each channel, SM backgrounds partly arising because of the limited detector acceptance, will be discussed.
To extract sensitivity contours in the model parameter space, the ultimate Belle II luminosity goal of 50~ab$^{-1}$ is assumed, and DP masses are scanned in the MeV to few GeV range.

\subsection{Monophoton}
In this section we consider the monophoton signature consisting of a single observed photon together with missing transverse energy, i.e.\ $e^+e^- \to \gamma + \slashed{E}$, see e.g.\ Ref.~\cite{DePietro:2018sgg}.
The DP contributes to this signature as missing energy when it decays into an invisible final state or outside of the detector acceptance.
Its dominant contribution to this signature stems from the decays into $\mu$ and $\tau$ neutrinos, $X\to \nu_\alpha \bar \nu_\alpha,$ with $\alpha = \mu,\tau$, which have a large ${\cal O}(1)$ branching ratio, while the fraction of decays outside the detector phase space contributes always below one percent to the total DP signal and will be neglected.

\begin{figure}
    \centering
    \includegraphics[width=0.8\textwidth]{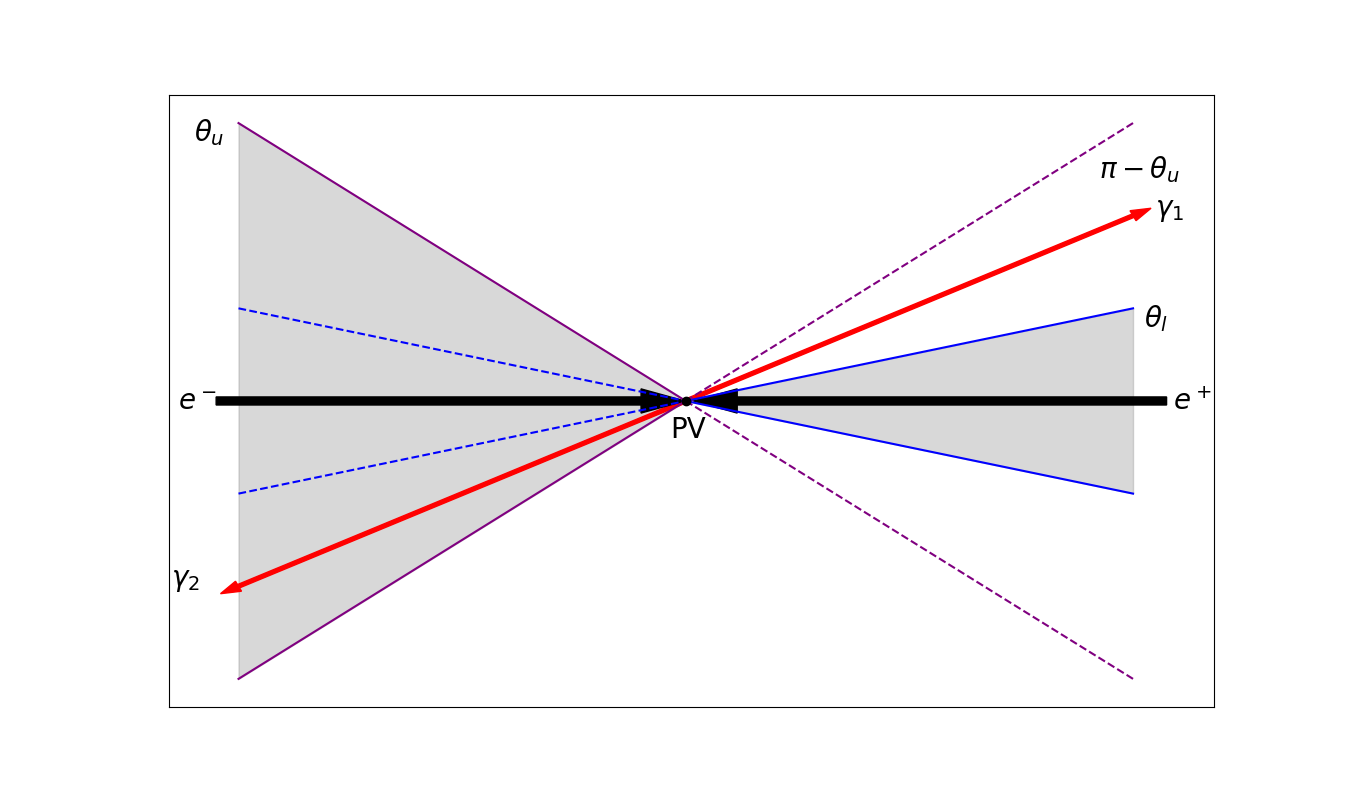}
    {\vspace{-0.6cm}}
    \caption{Schematic picture in the lab frame of a monophoton background event arising from the di-photon process $e^+e^- \to \gamma\gamma$ due to the detector geometry.}
    \label{fig:mono-photon_bkg}
\end{figure}

The most relevant SM background processes contributing to the monophoton signature are: di-photon processes ($e^+e^- \to \gamma \gamma$) where one photon escapes detection; radiative Bhabha scattering ($e^+e^- \to \gamma e^+ e^-$) where the final state leptons are undetected; and processes with a photon and one off-shell $Z$-boson decaying invisibly ($e^+e^- \to Z \gamma$ with $Z \to \nu\bar{\nu}$).
The cross sections of these SM processes representing the background for our following analyses, are given in the first group of Tab.~\ref{tab:SM_background_processes}, and they are obtained applying the following acceptance cuts on the final state particles: for the monophoton signature, we ask just one photon in falling within the geometric acceptance of the CDC and with a minimum energy cut at 1~GeV to suppress various backgrounds~\cite{DePietro:2018sgg, Belle-II:2023cal}, and for the di-photon background we require the additional photon to fall outside the detector acceptance; for the two muons plus missing energy and four muons signatures we ask for two and four muons within the detector acceptance respectively.
The large di-photon and the radiative Bhabha production cross sections also lead to a large number of monophoton events due to the gap in acceptance coverage between the forward and backward hemispheres of the Belle II detector.
Monophoton signatures from these final states occur when one photon, or the lepton pair, is missed. Such a scenario is depicted in Fig.~\ref{fig:mono-photon_bkg}, where one photon, $\gamma_1$, is in the forward direction with an angle larger than $\theta_l=17^\circ$, while the second photon, $\gamma_2$, escapes detection in the backward direction with an angle larger than $\theta_u=150^\circ$.
The background contribution from the electro-weak process with a $Z$-boson decaying into neutrinos was not considered in previous analyses~\cite{DePietro:2018sgg} because of its comparatively small cross section.
Nevertheless, we expect a significant contribution from this SM background process because of the high target luminosity of the Belle II experiment, and we therefore include this contribution in our analysis.

We simulate event samples at tree-level containing $10^6$ events scanning over 35 DP masses in the range 1~MeV$\leq m_X \leq$ 8~GeV, $10^6$ events for the SM di-photon and off-shell $Z$ into neutrinos processes, and $10^5$ events for the radiative Bhabha process.
Di-photon and radiative Bhabha samples are produced with no angular acceptance cuts on the final state particles.
We extract their contributions to the monophoton SM background selecting the events where respectively one of the photons or the electron-positron pair are outside of the detector acceptance. 
To the di-photon sample, we apply the following selection condition:
\begin{equation}
  (\theta_l < \theta_{\gamma_i} < \theta_u) \quad \text{and} \quad (\theta_{\gamma_j} < \theta_l \text{ or } \theta_{\gamma_j} > \theta_u)\,,\quad \text{with}\quad i\neq j \quad \text{and} \quad i,j=1,2\,,
\end{equation}
where $\theta_{\gamma_i}$ is the angle of photon $\gamma_i$ with respect to the positive beam axis, and $\theta_l = 17^\circ$ and $\theta_u = 150^\circ$.
The surviving monophoton events amount to just 1\% of the original di-photon sample. They populate a characteristic phase space region, with photons being highly energetic $E_\gamma \approx$ 7~GeV and in a narrow forward region $\theta_l < \theta_\gamma < \pi - \theta_u$ or photons with lower energy $E_\gamma \approx$ 4~GeV confined to a narrow backward region $\theta_u - \theta_l < \theta_\gamma < \theta_u$.
Similarly, from the radiative Bhabha sample, we select monophoton events with both electron and positron having angles smaller than $\theta_l$ or bigger than $\theta_u$.
This acceptance cut is very effective and reduces this sample by a factor of $10^{-5}$.
The off-shell $Z$-boson sample constitutes an irreducible background for the DP signal in this channel, as its topology is identical to the DP signal, despite the $Z$-boson being much heavier than the DP in this model.

\begin{table}
\centering
\begin{tabular}{ccccccc}
\toprule
 $m_X$ [MeV] & $E_\gamma$ [GeV] & $\epsilon_{\gamma\nu\bar{\nu}} (\%)$  & $\epsilon_X (\%)$& $N_{\rm bkg}$ & $\sigma_X^{\rm crit}$ [fb]\\
\hline
1 &
[4.10, 6.90] &
0.0747  &  92.18  & 88 & 0.00035 \\  
10 &
[4.10, 6.90] &
0.0747  &  92.19  & 88 & 0.00035 \\  
100 &
[4.10, 6.90] &
0.0748  &  92.20  & 89 & 0.00035 \\  
200 &
[4.10, 6.90] &
0.0753  &  92.20  & 89 & 0.00036 \\  
300 &
[4.09, 6.89] &
0.0769  &  92.13  & 91 & 0.00036 \\  
500 &
[4.09, 6.89] &
0.0802  &  92.20  & 95 & 0.00038 \\  
1000 &
[4.06, 6.51] &
0.0980  &  92.20  & 116 & 0.00042 \\  
2000 &
[3.95, 6.40] &
0.2116  &  92.38  & 250 & 0.00060 \\  
3000 &
[3.77, 6.22] &
0.4903  &  92.53  & 580 & 0.00089 \\  
4000 &
[3.51, 5.61] &
0.9472  &  92.77  & 1120 & 0.00122 \\  
5000 &
[3.18, 5.28] &
1.7943  &  93.06  & 2122 & 0.00166 \\  
6000 &
[2.78, 4.53] &
3.0736  &  93.34  & 3635 & 0.00216 \\  
7000 &
[2.30, 3.70] &
5.4986  &  93.58  & 6502 & 0.00287 \\  
8000 &
[1.76, 2.81] &
10.2567  &  93.83  & 12129 & 0.00389 \\  
\hline  

\end{tabular}
\caption{Details of the monophoton analysis for few representative DP masses.
Column $E_\gamma$ shows the considered interval of photon energy, as discussed in the text.
Columns $\epsilon_{\gamma\nu\bar{\nu}}$ and $\epsilon_X$ give the efficiencies of the optimal cuts on the $\gamma\bar\nu\nu$ background and on the signal, respectively.
Column $N_{\rm bkg}$ contains the total number of $\gamma\nu\bar{\nu}$ background events in the signal regions defined by the optimal cuts described in the text.
Column $\sigma_X^{\rm crit}$ shows the critical cross section for a signal visible at 95\% CL for an integrated luminosity of 50~ab$^{-1}$.}
\label{tab:monophoton}
\end{table}

\begin{figure}
    \centering
    \includegraphics[width=0.8\textwidth]{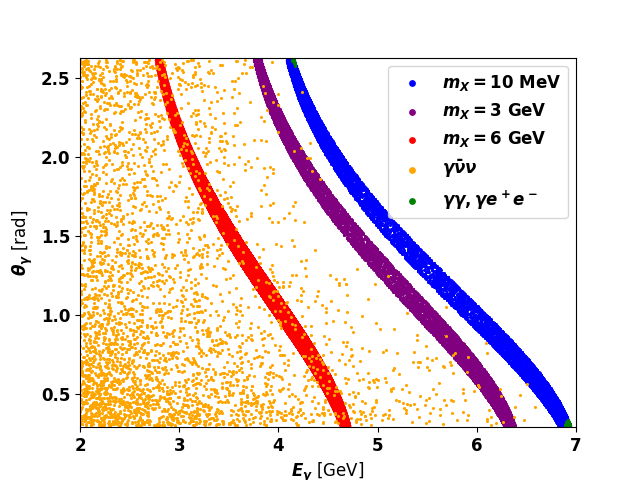}
    \caption{Distribution in angle and energy of the simulated monophoton sample events in the full Belle II detector fiducial volume, from DP signals with $m_X$ = 10~MeV, $m_X$ = 3.0~GeV and $m_X$ = 6.0~GeV (blue, purple and red respectively), from SM QED di-photon and radiative Bhabha processes (green), and for the irreducible EW $Z$ to neutrinos process (yellow).
    }
    \label{fig:monophoton_E_Theta}
\end{figure}

In Fig.~\ref{fig:monophoton_E_Theta} we show the distribution in angle and energy for the three SM background samples, and for three representative signal samples corresponding to $m_X=10$~MeV, $m_X=3.0$~GeV and $m_X=6.0$~GeV.
In the DP signal samples, the photon's angle and energy are tightly correlated and all the signal events are confined to a narrow band in the $\theta_\gamma$-$E_\gamma$ plane.
The background events from di-photon and radiative Bhabha samples (green points) populate a defined phase space region as discussed above, and can be entirely removed with a simple selection cut $20^\circ < \theta_\gamma < 148^\circ$.
This cut has selection efficiencies on the signal $\epsilon_S = 0.92 - 0.94$, across the DP mass range, and $\epsilon_{\gamma\nu\bar{\nu}} = 0.93$ on the sample with a photon and one off-shell $Z$-boson.
In the latter background process, the heavy $Z$-boson takes most of the energy and the photon is preferentially soft, therefore in the sample we observe no clear correlation between photon angle and energy.

We determine the sensitivity to the DP signal defining optimal cuts, which exploit the kinematic features of signal and background, in the following way.
For each of the 35 DP masses, we consider bins of photon energies $E_\gamma$ between 2~GeV and 7~GeV and width $\Delta E$ = 0.35~GeV, compatible with the detector energy resolution of 5\% for $E\sim 7$~GeV~\cite{Belle-II:2018jsg}.
For each $E_\gamma$ bin we identify the interval in $\theta_\gamma$ that encloses the whole signal sample, obtaining bins of width $\Delta \theta \approx$ 0.5~rad.
This procedure defines, for each DP mass, optimal search regions in the kinematic plane, which follow the signal distribution.
For DP masses $m_X \leq 4$~GeV, this selection suppresses the $\gamma\bar\nu\nu$ background by at least a factor ${\cal O}(10^{-3})$.
For heavier DPs, the kinematic distribution of signal events shifts to lower energies, where there is a higher background event rate, as visible in Fig.~\ref{fig:monophoton_E_Theta}.
With the target integrated luminosity of 50~ab$^{-1}$, the individual optimal search regions contain at most about $10^3$ $\gamma\bar\nu\nu$ background events for $m_X \leq 4$~GeV, rising to $10^4$ events for $m_X$ = 8~GeV.
Tab.~\ref{tab:monophoton} summarises the main analysis results for a few selected DP masses.
The second column shows the overall $E_\gamma$ interval, which spans all the sub-bins with width $\Delta E_\gamma$, as discussed above, and which encloses the whole signal sample for each DP mass.
The bins in the angle $\theta_\gamma$ span the whole detector acceptance.
In addition an angular cut $20^\circ < \theta_\gamma < 148^\circ$ is applied to suppress the di-photon and radiative Bhabha backgrounds.
The total efficiencies after our optimal cuts on background and signal are given in the third and fourth columns of Tab.~\ref{tab:monophoton}.
The fifth column shows the total $\gamma\nu\bar{\nu}$ background event rate normalised to the integrated luminosity falling in the optimal search regions.
We calculate the critical cross section for a visible signal above 95\% CL for an integrated luminosity of 50~ab$^{-1}$ according to the procedure described in Sect.~\ref{sec:analysis_methodology}, including an additional factor $\epsilon_\gamma = 0.99$ to account for the reconstruction efficiency of photons with energy at or above 2~GeV~\cite{Henrikas:2604}.
The obtained values are reported in the last column of Tab.~\ref{tab:monophoton}.
The corresponding sensitivity contour is displayed in Fig.~\ref{fig:summary_plot} (dotted dark blue line) and discussed in Sect.~\ref{sec:results}.

\subsection{Two muons plus missing energy}
We now discuss the signature of two muons with missing transverse energy, $e^+e^- \to \mu^+\mu^- + \slashed{E}$.
The signal in this channel originates from a DP, radiated from a muon and then decaying invisibly into muon or tau neutrinos: $e^+e^- \to \mu^+\mu^- + X \to \mu^+\mu^-\nu_{\mu,\tau}\bar{\nu}_{\mu,\tau}$.
The contribution from DPs decaying outside the detector acceptance is at most at the sub-percent level for most of the considered parameter space, and will therefore be neglected.
Three SM background processes contribute to this signature, and are listed in the second group of Tab.~\ref{tab:SM_background_processes} with their cross sections.

The first arises from di-tau production where both taus decay into muons, $e^+e^- \to \tau^+\tau^-$ and $\tau^\pm \to \mu^\pm \nu_\mu \bar\nu_\tau$, with cross section $\sigma_{\tau\tau\to \mu\mu} = 0.8~{\rm nb} \times {\rm Br}(\tau \to \mu)^2\times \epsilon_{\mu\mu}$, where ${\rm Br}(\tau \to \mu) = 17.39\%$ is the branching ratio of the tau lepton into muons~\cite{ParticleDataGroup:2022pth}. An efficiency of $\epsilon_{\mu\mu} = 0.87$ for this background, due to the geometric acceptance for the two muons, has been determined from simulation.
The second SM background is di-muon production with a hard photon radiated outside the detector acceptance region, $e^+e^- \to \mu^+\mu^-\gamma_{\rm inv}$, with a cross section of $\sigma_{\mu\mu\gamma} = 0.32$~nb. 
The third SM background is di-muon production associated with a di-electron pair falling outside the detector acceptance.
Imposing this condition on the electron pair completely suppresses this background to less than one event for the target integrated luminosity and therefore it can be safely neglected.

We simulate samples with $10^{6}$ events for each background process, and for 36 DP masses in the range 1~MeV$ \leq m_X \leq $~9~GeV.
The following observables are constructed from the muons' four-momenta: di-muon energy $E_{\mu\mu}$, three momentum $P_{\mu\mu}$, and angle $\theta_{\mu\mu}$ with respect to the electron beam direction, and the di-muon opening angle $\alpha_{\mu\mu}$.
In addition we also construct the recoil mass of the di-muons, which is defined as
\begin{equation}
 m_{\rm recoil}^2 \ =\ \left(E_{e^-} - E_{\mu\mu}\right)^2  - \Delta E^2 - P_{\mu\mu}^2 + 2\Delta E\, \cos(\theta_{\mu\mu}) P_{\mu\mu}\,,
\end{equation}
where $E_{e^-}=11$~GeV is the electron beam energy, and $\Delta E = 3$~GeV is the difference between electron and positron beam energies.
The recoil mass peaks at $m_X^2$ for the DP signal, at zero for the radiative di-muon background, and it is smeared from 0 to $s$ for the di-tau background due to the energy carried away by the neutrinos from the tau decays.
Optimal cuts in the observables $\alpha_{\mu\mu}$, $\theta_{\mu\mu}$ and $P_{\mu\mu}$ were found to improve the sensitivity only marginally, and will hence not be discussed in the following.
\begin{figure}
    \centering
    \includegraphics[width=0.48\textwidth]{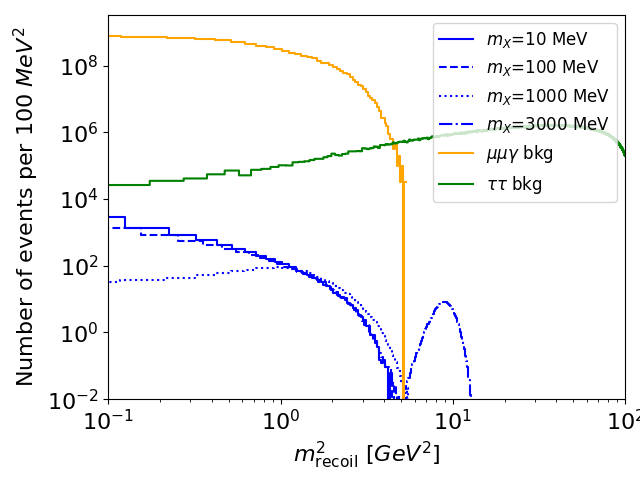}
    \includegraphics[width=0.48\textwidth]{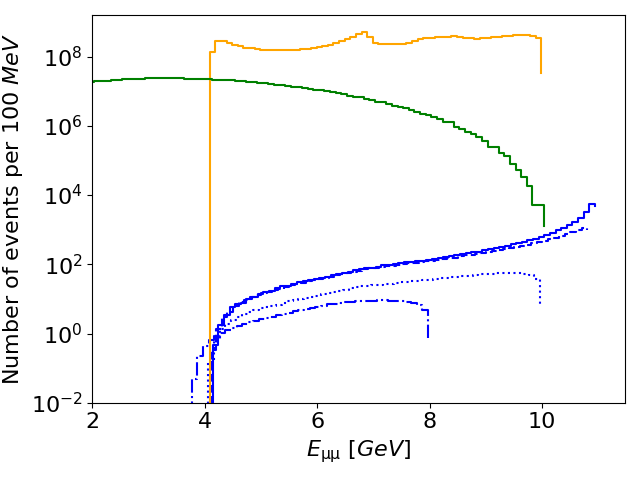}
    {\vspace{-0.4cm}}
    \caption{Number of event for an integrated luminosity of 50~ab$^{-1}$ as function of the recoil mass $m^2_{\textrm{recoil}}$ (left) and of the di-muon energy $E_{\mu\mu}$ (right), normalised as indicated in the axes captions. The blue signal curves correspond to the choices $m_X$ = 10, 100, 1000 and 3000~MeV and $g_X = 0.0007$. The green curves denote the di-tau background and include the decay Br($\tau \to\mu$). The yellow curves denote the radiative di-muon background for this signature.}
    \label{fig:dimuon-observables}
\end{figure}
We find that the two most relevant observables for signal-over-background enhancement are the recoil mass $m_{\rm recoil}^2$ and $E_{\mu\mu}$, whose distributions are shown in Fig.~\ref{fig:dimuon-observables} for DP masses of $m_X = 10,\,100,\,1000,\,3000$~MeV and $g_X = 0.0007$.
Given the shape of the recoil mass distribution, we find that appropriate cuts in this observable can almost completely remove contributions from the di-muon background for $m_X \geq 1$~GeV, in agreement with the results of Ref.~\cite{DePietro:2018sgg}.
For DPs with $m_X \leq 1$~GeV, the most effective way to disentangle the signal from the background is through optimal cuts in the $E_{\mu\mu}$ observable.
In our analysis, we use optimal cuts in both observables for all $m_X$ to derive the sensitivity.
In Tab.~\ref{tab:dimuon_missingE}, for some representative DP masses, we list the selection cuts and their efficiencies on the two SM backgrounds and on the signal.
The critical signal cross sections in the last column are obtained following Eq.~\eqref{eq:limit-gauss} assuming an integrated luminosity of 50~ab$^{-1}$, and they translate in the 95\% CL exclusion solid dark blue contour line in Fig.~\ref{fig:summary_plot}.

\begin{table}
\centering
\begin{tabular}{cccccccc}
\toprule
 $m_X$ [MeV] &  $m_{\rm recoil}^2$ [GeV$^2$] &  $E_{\mu\mu}$ [GeV] &  $\epsilon_{\mu\mu\gamma} (\%)$ & $\epsilon_{\tau\tau} (\%)$ & $\epsilon_X (\%)$ & $\sigma_X^{\rm crit}$ [fb]\\
\hline
1  & 
[-4.2, 3.1] & 
[7.2, 11] & 
16.1  &  0.56  &  29.3  &  8.82 \\
10  & 
[-4.4, 3.4] & 
[5.2, 11] & 
43.7  &  0.73  &  99.7  &  4.27 \\
100  & 
[-4.4,-0.1] & 
[7.5, 11] & 
5.5  &  0.003  &  27.2  &  5.55 \\
200  & 
[-4.3,-0.3] & 
[7.4, 11] & 
3.1  &  0.001  &  18.9  &  6.00 \\
300  & 
[0.08, 4.1] & 
[7.4, 11] & 
5.1  &  0.85  &  39.6  &  3.68 \\
500  & 
[0.38, 4.6] & 
[7.9, 10] & 
1.3  &  0.91  &  23.5  &  3.18 \\
1000  & 
[0.25, 5.5] & 
[7.3, 10] & 
2.5  &  1.38  &  65.1  &  1.59 \\
2000  & 
[2.1, 8.3] & 
[6.7, 9.0] & 
0.1  &  2.65  &  71.6  &  0.349 \\
3000  & 
[5.9, 13] & 
[3.8, 8.0] & 
0.0  &  5.81  &  99.8  &  0.307 \\
4000  & 
[13, 20] & 
[3.5, 7.0] & 
0.0  &  7.33  &  99.7  &  0.345 \\
5000  & 
[22, 28] & 
[3.2, 6.0] & 
0.0  &  7.85  &  99.7  &  0.357 \\
6000  & 
[34, 39] & 
[2.8, 5.0] & 
0.0  &  7.61  &  99.6  &  0.352 \\
7000  & 
[47, 52] & 
[2.3, 3.9] & 
0.0  &  6.29  &  92.2  &  0.346 \\
8000  & 
[62, 67] & 
[1.8, 2.9] & 
0.0  &  4.51  &  86.7  &  0.312 \\
9000  & 
[80, 83] & 
[1.9, 1.9] & 
0.0  &  0.05  &  21.6  &  0.135 \\
\hline                  
\end{tabular}
\caption{Selection cuts in the di-muon plus missing energy channel for some representative DP masses in the two observables: recoil mass $m_{\rm recoil}^2$ and di-muon energy $E_{\mu\mu}$. 
Columns labelled $\epsilon_{\mu\mu\gamma}$ and $\epsilon_{\tau\tau}$ denote the selection efficiency for the SM background processes with $\mu\mu\gamma$ and $\tau\tau$ final states respectively.
Column $\epsilon_X$ contains the selection efficiency on the signal, and the last column, $\sigma_X^{\rm crit}$, shows the critical cross section for a signal visible at 95\% CL for an integrated luminosity of 50~ab$^{-1}$.
}
\label{tab:dimuon_missingE}
\end{table}

\subsection{Four muons prompt}
Here we study the sensitivity of the Belle II experiment on the DP signal in the final state with two muon pairs, $e^+ e^- \rightarrow \mu^+ \mu^+ \mu^- \mu^-$, which we will refer to as the $4\mu$ channel in the following.
The DP signal in this channel stems from a DP radiated from a muon and decaying into a muon pair, see the left panel of Fig.~\ref{fig:feynman}.
The main SM background is from the QED process with two photons decaying into muon pairs and has a cross section of $\sigma_{4\mu} = 155$~fb (see Tab.~\ref{tab:SM_background_processes}).
Here we neglect additional possible backgrounds with hadronic final states where pions are misidentified as muons, which has a probability of 1.4\% per pion only.

\begin{figure}
\centering
\includegraphics[width=0.8\textwidth]{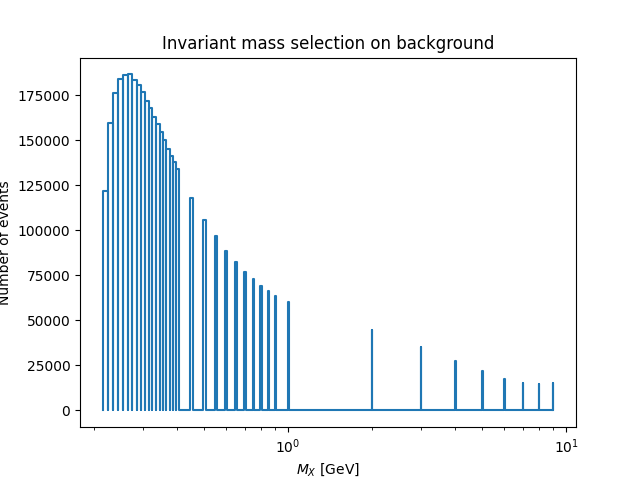}
{\vspace{-0.4cm}}
\caption{Distribution of expected number of background events for an integrated luminosity of 50~ab$^{-1}$ with the full bin scheme adopted in our analysis. Explicit values for some representative bins are given in Tab.~\ref{tab:prompt-four-muon}.
The events in each bin represent the background for each DP mass search region.}
\label{fig:inv_mass_sel_eff}
\end{figure}

We simulate 10$^6$ events within the geometric acceptance for the background, and $10^6$ signal events for 32 DP masses in the range 220~MeV~$\leq m_X\leq$~9~GeV. In the signal samples, the DP propagator is taken to be on-shell as the DP is very narrow. The DP kinematics is then reconstructed by identifying the muon pair with an invariant mass $m_X\pm 5$~MeV, thus taking into account the finite detector resolution. With this procedure, the signal selection efficiency is basically 100\% in all bins. Due to the tight selection, the resulting background suppression ranges from around 2\% in bins with $m_X$ close to the di-muon threshold, to $\sim$ 0.2\% for $m_X \simeq \sqrt{s}$.
This is visible in Fig.~\ref{fig:inv_mass_sel_eff} where the number of background events within each search window for an integrated luminosity of 50~ab$^{-1}$ is shown.

\begin{figure}
    \centering
    \includegraphics[width=0.48\textwidth]{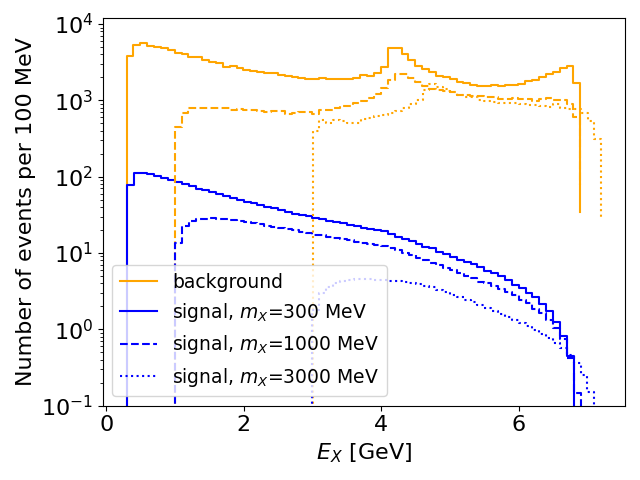}
    \includegraphics[width=0.48\textwidth]{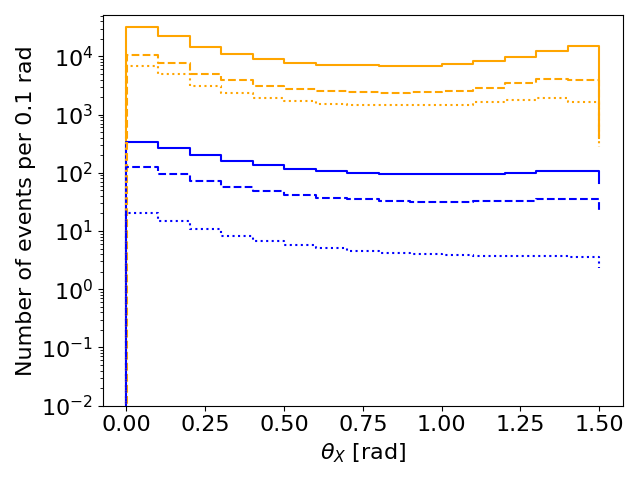}
    {\vspace{-0.4cm}}
    \caption{Energy and angular distributions of dark photon signals (blue) and SM background (yellow) in the 4$\mu$ final state at Belle II with integrated luminosity of 50~ab$^{-1}$. Signal curves are shown for $m_X$ = 300, 1000 and 3000~MeV.}
    \label{fig:fourmuon-kinematics}
\end{figure}
\begin{table}
    \centering
    \begin{tabular}{c c c c c c c c}
        \toprule
       $m_X$ [MeV] & $m_{\mu\mu}$ [GeV] & $\epsilon_{m_{\mu\mu}} (\%)$  & $E_X$ [GeV]  & $\theta_X$ [rad]& $\epsilon_{4\mu} (\%)$ & $\epsilon_X (\%)$ & $\sigma_X^{\rm crit}$ [fb] \\
       \hline
220 &
[0.215, 0.225] &
1.57 &
[0.22, 3.88] &
[0.00007  ,  2.94] &
65.09 & 90.35 & 0.0101 \\ 
300 &
[0.295, 0.305] &
2.28 &
[0.30, 3.92] &
[0.00074  ,  2.97] &
61.09 & 89.31 & 0.0119 \\   
500 &
[0.495, 0.505] &
1.37 &
[0.50, 3.70] &
[0.0031  ,  2.98] &
49.13 & 83.79 & 0.0088 \\  
1000 &
[0.995, 1.005] &
0.78 &
[1.00, 3.67] &
[0.54  ,  3.14] &
32.80 & 75.42 & 0.0060 \\ 
2000 &
[1.995, 2.005] &
0.57 &
[2.00, 4.02] &
[1.04  ,  2.72] &
22.95 & 62.36 & 0.0052 \\ 
3000 &
[2.995, 3.005] &
0.46 &
[3.00, 4.69] &
[1.38  ,  2.79] &
28.43 & 60.59 & 0.0054 \\ 
4000 &
[3.995, 4.005] &
0.35 &
[4.00, 6.26] &
[1.38  ,  3.05] &
70.72 & 88.79 & 0.0051 \\ 
5000 &
[4.995, 5.005] &
0.28 &
[5.00, 6.97] &
[1.60  ,  3.14] &
78.48 & 91.86 & 0.0046 \\ 
6000 &
[5.995, 6.005] &
0.23 &
[6.00, 7.67] &
[1.80  ,  3.14] &
83.36 & 93.74 & 0.0042 \\ 
7000 &
[6.995, 7.005] &
0.20 &
[7.00, 8.44] &
[1.95  ,  3.08] &
90.61 & 95.24 & 0.0040 \\ 
8000 &
[7.995, 8.005] &
0.19 &
[8.00, 9.25] &
[2.09  ,  3.14] &
100.00 & 100.00 & 0.0039 \\ 
9000 &
[8.995, 9.005] &
0.19 &
[9.08, 9.86] &
[2.30  ,  3.10] &
99.86 & 99.57 & 0.0040 \\ 
\hline
    \end{tabular}
    \caption{Selection cuts for the 4$\mu$ channel for the some representative DP masses.
    Column $m_{\mu\mu}$ contains the invariant mass selection cuts, and $\epsilon_{m_{\mu\mu}}$ the corresponding fraction of background events in that interval.
    Columns $E_X$ and $\theta_X$ give the selection cuts on the two observables.
    Columns $\epsilon_{4\mu}$ and $\epsilon_X$ contain the efficiency of the combined cuts on the background and signal respectively.
    The last column, $\sigma_X^{\rm crit}$, shows the critical cross section for a signal visible at 95\% CL for an integrated luminosity of 50~ab$^{-1}$.
    }
    \label{tab:prompt-four-muon}
\end{table}

To achieve the best sensitivity for DP signals in this channel, we consider four observables for the DP candidates: energy $E_X$, momentum, angle $\theta_X$ with respect to the beam axis, and opening angle of the two additional muons.
The distributions for $E_X$ and $\theta_X$ are shown in Fig.~\ref{fig:fourmuon-kinematics} for the background and for signals from DP masses $m_X$ = 300, 1000 and 3000~MeV.
For each DP mass used in the scan, optimal cuts on the four kinematic observables are determined to maximise the 95\% CL signal exclusion, following Eq.~\eqref{eq:limit-gauss}. 
We find that cuts on $E_X$ and $\theta_X$ lead to the best signal over background enhancement. 
A list of the optimised cuts for some representative DP masses is given in Tab.~\ref{tab:prompt-four-muon}, together with the corresponding signal and background efficiencies due to the optimised cuts but after the initial invariant mass selection\footnote{Even for small values of $\theta_X$, the final state muons are all within the detector acceptance.}.
The last column gives the critical signal cross sections for an integrated luminosity of 50~ab$^{-1}$, from which we obtain the 95\% CL exclusion dashed-dotted dark blue contour line in Fig.~\ref{fig:summary_plot}, following the procedure described in Sect.~\ref{sec:analysis_methodology}.

\subsection{Displaced vertices}
\begin{figure}
    \centering
    \includegraphics[width=0.8\textwidth]{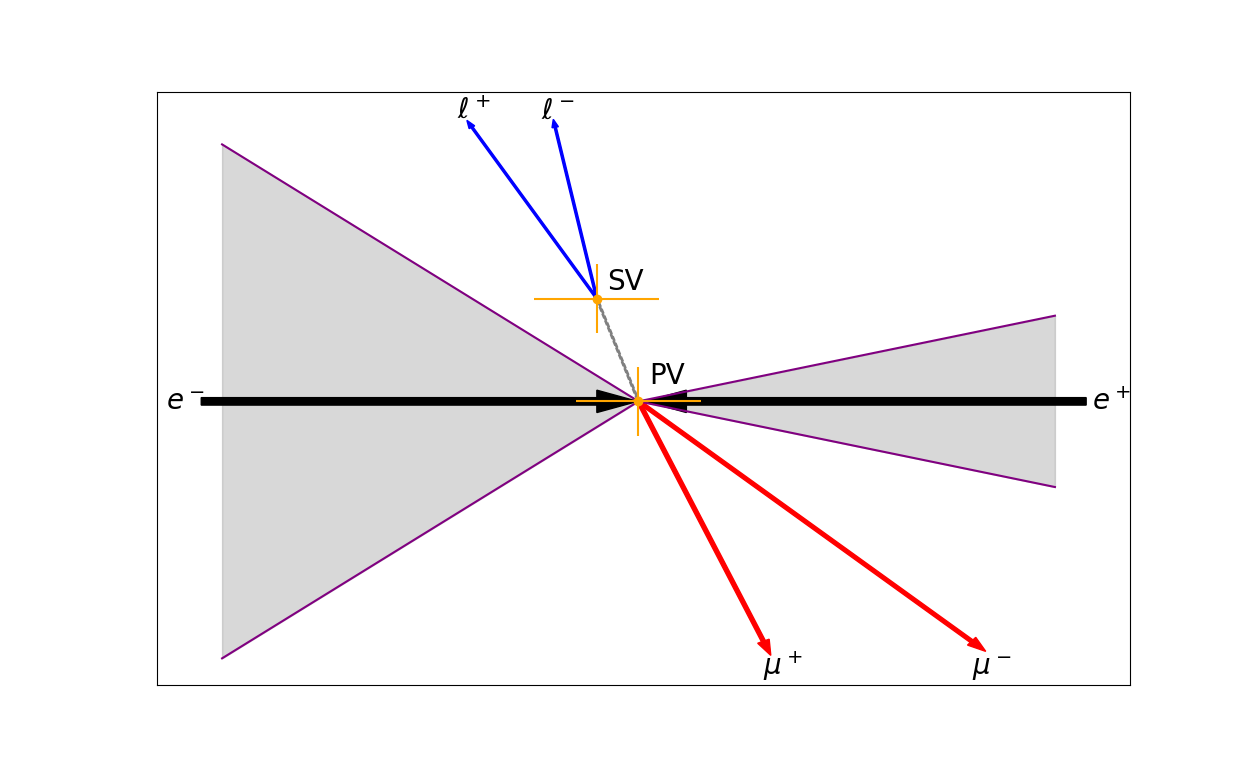}
    {\vspace{-0.6cm}}
    \caption{Schematic description of the process $e^+e^-\to \mu^+\mu^-X$, where $X$ propagates a macroscopic distance and decays into $\ell^+\ell^-$. The primary vertex is denoted by `PV' and is identified with the intersection of the two muon tracks (red arrows), the secondary vertex is denoted with `SV' and identified with the intersection of the two lepton tracks (blue arrows). Uncertainties in the PV and SV positions are indicated by the yellow error bars.
    The electron and positron beams are denoted by black arrows, the detector acceptance is indicated by the purple lines, where the grey regions are invisible.}
    \label{fig:displacedvertex_schematic}
\end{figure}
We consider a special signature in the 4$\ell$ final state which features two interaction vertices separated by a measurable distance.
Experimentally the position of the decay vertices is reconstructed from the information collected in the Vertex detector (VXD), which is the sub-detector closest to the interaction point~\cite{Belle-IISVD:2017cpj}, and whose main purpose is to study the decay of tau leptons and $B$ and $D$ mesons.
The design and the performance of the VXD~\cite{Belle-IISVD:2022upf, Belle-II:2020mvd} allow a displacement resolution in the transverse plane as precise as a few micrometers~\cite{Wieduwilt:2021phh}. 
A displaced signature is observed when the distance between the two vertices exceeds the detector resolution. This is schematically depicted in Fig.~\ref{fig:displacedvertex_schematic}, where the Primary Vertex (PV) is on the nominal beam axis while the Secondary Vertex (SV) is displaced.\footnote{The position of the vertices along the beam direction is reconstructed less precisely than that in the transverse plane. For our analysis, we will therefore only consider the transverse displacement. We have checked that including information on the longitudinal displacement, even under optimistic assumptions, adds only marginally to the derived sensitivities.}
In our analysis for displaced vertices, we consider a spatial detector resolution of 5~$\mu$m, which is somewhat optimistic compared to the recent study~\cite{Maiti:2023axl}, but potentially achievable in the future as dedicated studies on the VXD performance are ongoing~\cite{Belle-IISVD:2024gjd}.

In the $L_\mu - L_\tau$ model considered here, a light DP with mass $\sim$ 250~MeV and gauge coupling $g_X \sim 10^{-4}$ has a decay length in its rest frame $\lambda_X \sim 1~\mu$m (see Fig.~\ref{fig:Branching_ratio}).
The additional boost into the Belle II laboratory frame enhances the DP decay length up to a factor 10. Therefore, with the assumed vertex resolution and the high integrated luminosity collected at Belle II, a statistically significant number of DP decays populating the tail of the decay distribution will appear as events with a displaced vertex.
In principle, DP decays into any charged lepton contribute, but decays into electrons are strongly suppressed by the factor $\epsilon^2$ (cf. Eq.~\eqref{eq:loop-mixing}) compared to muons, whereas decays into taus require $m_X > 2m_\tau \simeq 3.5$~GeV, where the DP lifetime is small. We will therefore focus on displaced signatures with four muons.

SM background contributions with displaced final states include intermediate particles with a finite life-time such as $B$ mesons and $\tau$ leptons.
However, the Belle II detector is designed to perform precision measurements on the former, which have a very different topology compared to the DP signals considered, since on average $B$ decays yield many tracks and include several mesons.
Since the likelihood to misidentify several pions (or kaons) as muons is tiny, we assume that this background is sufficiently suppressed by the requirement of an exclusive four muon final state. 
The SM process $e^+e^- \to \tau^+\tau^- \mu^+\mu^-$ with $\tau^\pm \to \mu^\pm \bar{\nu}_\mu \nu_\tau$ has a cross section of about 4~fb and can feature displaced vertices because of the finite decay length of the relativistic $\tau$ lepton, $\lambda_\tau \simeq 10^{-4}$~m.
This background can be effectively vetoed by requiring the total deposited energy to be within 3\% of $\sqrt{s}$, which reduces the predicted number of such events to 0.1 for an integrated luminosity of 50~ab$^{-1}$.

The only relevant background for our analysis originates from the SM QED process with four muons in the final state where, because of the finite detector resolution, a virtual photon decay may appear with a visible displacement.
We generate $10^7$ QED background events, and $10^6$ signal events for 19 DP masses in the range $m_X$ = [220, 400]~MeV.
Since heavy DPs have on average a shorter decay length, in our model we lose sensitivity at higher masses, where the decay length is too short for the discrimination of the displaced vertex.
Similarly to the four muon prompt analysis, in each event we identify the muon and anti-muon pair which reconstruct the DP invariant mass with tolerance $m_X\pm 5$~MeV, obtaining the event distribution already shown in Fig.~\ref{fig:inv_mass_sel_eff}.

\begin{table}
    \centering
    \begin{tabular}{c c c  c c  c}
        \toprule
       $m_X$ [MeV] & $m_{\mu\mu}$ [GeV] & $\epsilon_{m_{\mu\mu}} (\%)$ & BFT [$\mu$m] & $\epsilon_X (\%)$ & $g_{X}$ \\
       \hline
 220 &
 [0.215, 0.225] &
1.6 &
15.18 & [44.19, ~7.13] &
[0.63, 1.6] $\times$ 10$^{-4}$ \\
230 &
 [0.225, 0.235] &
2.1 &
13.86 & [44.10, ~7.55] &
[0.63, 1.6]  $\times$ 10$^{-4}$ \\
240 &
 [0.235, 0.245] &
2.3 &
13.99 & [42.96, ~6.10] &
[0.63, 1.6]  $\times$ 10$^{-4}$ \\
250 &
 [0.245, 0.255] &
2.4 &
14.96 & [36.69, 16.70] &
[0.63, 1.0]  $\times$ 10$^{-4}$ \\
260 &
 [0.255, 0.265] &
2.4 &
13.82 & [37.75, ~4.49] &
[0.63, 1.0]  $\times$ 10$^{-4}$ \\
270 &
 [0.265, 0.275] &
2.4 &
13.72 & [36.67, 16.27] &
[0.63, 1.0]  $\times$ 10$^{-4}$ \\
280 &
 [0.275, 0.285] &
2.4 &
13.91 & [34.11, 14.53] &
[0.63, 1.0]  $\times$ 10$^{-4}$ \\
290 &
 [0.285, 0.295] &
2.3 &
14.20 & [32.26, 12.66] &
[0.63, 1.0]  $\times$ 10$^{-4}$  \\
300 &
 [0.295, 0.305] &
2.3 &
13.53 & [31.71, 12.84] &
[0.63, 1.0]  $\times$ 10$^{-4}$ \\
310 &
 [0.305, 0.315] &
2.2 &
12.17 & [34.98, 14.97] &
[0.63, 1.0]  $\times$ 10$^{-4}$ \\
320 &
 [0.315, 0.325] &
2.2 &
13.78 & [28.18, 13.49] &
[0.63, 1.0]  $\times$ 10$^{-4}$ \\
\hline
    \end{tabular}
    \caption{Selection cuts for the 4$\mu$ displaced channel for the DP masses where sensitivity on the displaced signature is found.
    Column $m_{\mu\mu}$ contains the invariant mass selection cuts, and $\epsilon_{m_{\mu\mu}}$ the corresponding fraction of background events in that interval.
    Column BFT shows the background free thresholds above which no displaced background events are observed for a detector resolution of 5~$\mu$m and an integrated luminosity of 50~ab$^{-1}$.
    Column $\epsilon_X$ lists the efficiencies of the cuts on the signal including the BFT condition, returning at least 2.99 signal events for the target luminosity, and the corresponding ranges for the coupling $g_X$ given in the last column.
    }
    \label{tab:displaced-four-muon}
\end{table}

A random displaced position is assigned to each muon to every event in both background and signal samples.
The displacements of QED background events account for the finite detector resolution, and are generated drawing a random number for each event from a normal distribution with standard deviation equal to the considered detector resolution.
The displaced positions of the individual muons are assigned projecting the random Gaussian shift on the transverse $x$-$y$ plane according to the muons' individual momenta, i.e. multiplying the random Gaussian shifts by $p_{i}/ |\mathbf{p}|$, with $i = x,y$.\footnote{We chose to rescale the displacement according to the total momentum, rather than the transverse momentum, in order to account for the boost component along the beam axis.}
This procedure has been designed to resemble the experimental practice for the identification of particles' origins, which are derived from the particles' reconstructed tracks and momenta according to their energy deposition in the ECAL layers.
Effectively, with this procedure, the origin of each muon track lays in the direction of the particle momentum.
An additional displacement is added to the events in the signal samples, to account for the DP finite lifetime.
The additional displacements are assigned to the muon pair reconstructing the DP mass, and are calculated on an event-by-event basis to obtain the correct boost factor for the DP laboratory frame decay length $\lambda_X^\textrm{lab}$.
A random displacement for the pair is drawn following the DP exponential decay law, and then rescaled according to the muons' momenta and projected to the transverse plane to obtain the displacement of the individual muons.

After assigning a position to the origin of each muon as described, we select the events with visible displacements as follows.
In each event we calculate the distance between the muon and antimuon within either pair, and reject the ones where this is larger than the experimental resolution.
This selection removes events with wrongly paired muon-antimuon which effectively do not originate from the same vertex.
We then define the position of the PV and SV as the middle point between the paired muon and antimuon, and calculate the overall displacement as the relative distance between PV and SV.

We now determine, for each DP search region, the threshold distance above which the number of expected background events for an integrated luminosity of 50~ab$^{-1}$ drops below one, i.e.\ where the background is completely reducible and a background free analysis applies.
By construction, the determination of the end point of the distribution of background events suffers from large statistical fluctuations.
In order to reduce these fluctuations we repeat the assignment of random displacements to the background samples $10^4$ times, and then consider the average distribution of events as function of their displacement.
The end points of the distributions are obtained adopting a regression fitting method inspired by machine learning: we evaluate the number of surviving events above fixed multiples of the chosen resolution, and fit the logarithm of this distribution using a quadratic polynomial.
The parameters of the quadratic polynomial are obtained applying a linear regression algorithm, trained using a gradient descent minimiser with squared error cost function, and the optimised parameters are then used to estimate the distribution end point.
This procedure is repeated for each search region, to obtain the background free thresholds for each simulated DP mass, and gives more accurate results than a direct exponential fit of the background distributions.

In Tab.~\ref{tab:displaced-four-muon} the invariant mass selection and the corresponding efficiency on the background are listed for relevant DP masses together with the resulting background free thresholds (BFTs) for an integrated luminosity of 50~ab$^{-1}$ and a vertex resolution of 5~$\mu$m.
The last two columns of Tab.~\ref{tab:displaced-four-muon} show the range of efficiencies of the displaced selection on the signal and the corresponding range of $g_X$ couplings returning at least 2.99 surviving signal events for the target luminosity.
They are obtained repeating the same procedure applied to the background on the signal samples, estimating the surviving events above the BFTs and scanning over an appropriate range of the coupling $g_X$ which gives the boost factor and lab decay length to assign to the random draw for each event.
The analysis is sensitive only to DP masses between the di-muon production threshold and about 320~MeV, above which the decay length becomes too small.
In this range, the coupling $g_X$ is constrained between roughly 6~$\times$~10$^{-5}$ and 10$^{-4}$, as for smaller couplings the production cross section drops too low, while for higher couplings the decay length becomes too short.
The resulting 95\% CL exclusion contour is displayed in Fig.~\ref{fig:summary_plot} by the dashed dark blue line and will be further discussed in Sect.~\ref{sec:results}.

\subsection{Signatures with tau leptons}
Because of the underlying $L_\mu - L_\tau$ symmetry, processes involving the production of DPs in association with tau leptons, as in $e^+ e^- \to \tau^+\tau^- X$, have a cross section proportional to $g_X^2$, similar in magnitude to those discussed above involving muons (see Tab.~\ref{tab:signal_channels}, Tab~\ref{tab:SM_background_processes} and Fig.~\ref{fig:darkphoton_production_belle}).
Analogous search channels for DP detection can be defined replacing muons with taus in the final state: di-tau plus missing energy, di-taus plus two muons, four taus prompt, and two taus plus two displaced muons.
Tau leptons are not observable directly, but are detected reconstructing their decay products.
Their decays necessarily include light tau neutrinos which carry away a part of the kinematic information in form of missing energy or momentum.
Tau observables are therefore more `smeared out' compared to the analogous muon channels and have smaller selection and identification efficiencies, which eventually leads to an overall lower sensitivity.
Dedicated searches for light resonances decaying into tau pairs are being performed at Belle II~\cite{Belle-II:2023ydz}, which provide the existing most stringent limits in this signature.
Those constraints are however superseded by analogous analyses on signatures with muon final states.

For these reasons, in this study we do not further investigate the sensitivity of the Belle II detector to channels with tau leptons, but leave this for future work.
Indeed, if a positive signal would be discovered in any channel, signatures involving tau leptons could be decisive to fingerprint the underlying model.

\section{Results}
\label{sec:results}
\begin{figure}
    \centering
    \includegraphics[width=1.0\textwidth]{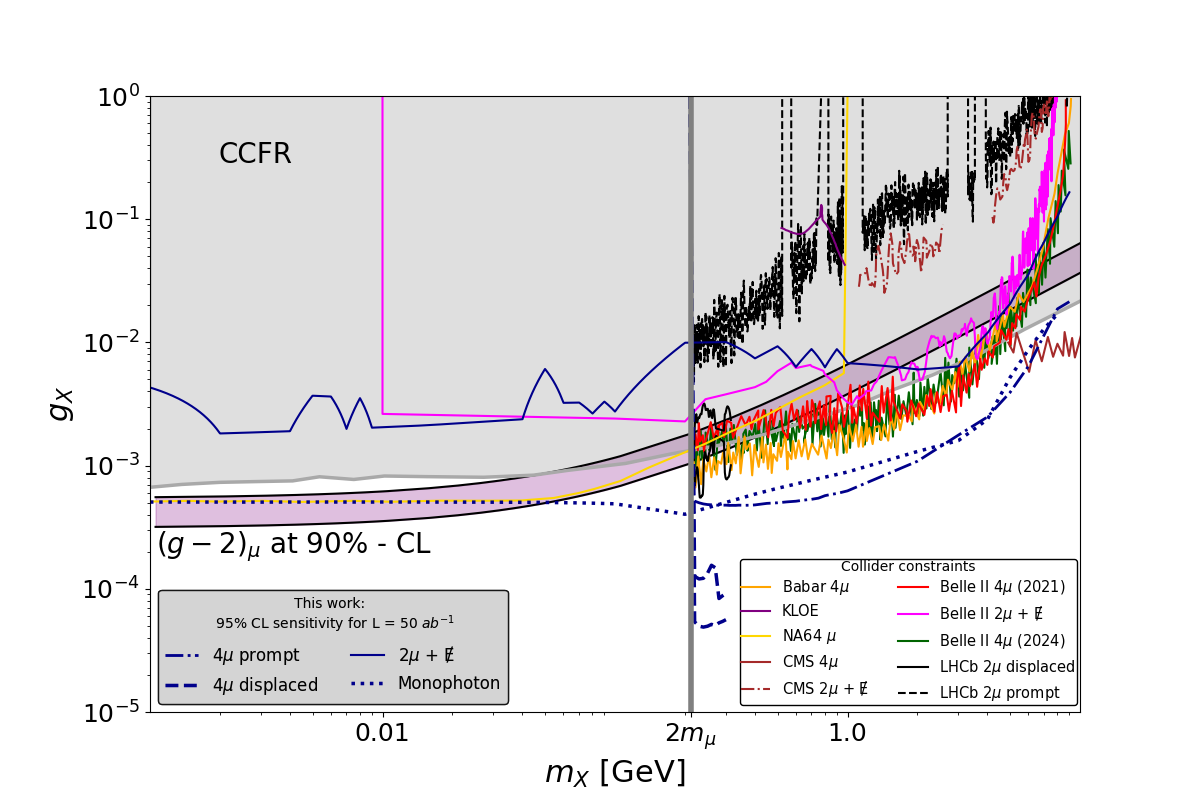}
    \caption{Summary of existing constraints and projected 95\% CL sensitivities for Belle II with integrated luminosity of 50~ab$^{-1}$. Constraints across the whole energy range are provided by the NA64 experiment with muon beam~\cite{NA64:2024klw} (solid yellow line), CCFR (solid light grey line) probing the neutrino trident production~\cite{Altmannshofer:2014pba} and by Belle II, probing the 2$\mu + \slashed{E}$ process (solid light pink line)~\cite{Belle-II:2022yaw}. Constraints that probe generic new light vector boson models are given by KLOE (solid purple line)~\cite{KLOE-2:2018kqf} and CMS (dashed brown line)~\cite{CMS:2023hwl}, both probing the 2$\mu + \slashed{E}$ final state, and by LHCb (dashed  black and solid black lines respectively)~\cite{LHCb:2019vmc} with a search in the prompt and displaced 2$\mu$ channels; with values of $\epsilon$ being recast into the $U(1)_{L_{\mu} - L_{\tau}}$ model considered here.  Other relevant constraints, in addition to CCFR, are given by Babar~\cite{BaBar:2016sci}, Belle II~\cite{Belle:2021feg,Belle-II:2024wtd} and CMS~\cite{CMS:2018yxg}, which look for resonances in the prompt 4$\mu$ channel. The purple band indicates the parameter range which would solve, at 90\% confidence level, the $4.2\sigma$ muon $g-2$ discrepancy, see the text for further details.}
    \label{fig:summary_plot}
\end{figure}

In this section we present our projections for the Belle II sensitivity on $L_\mu - L_\tau$ DPs in the search channels discussed above, assuming an integrated luminosity of 50~ab$^{-1}$.
The exclusion limits derived through the analyses described above are displayed in Fig.~\ref{fig:summary_plot} in the two dimensional parameter space of the model, scanning over the mass $m_X$ and coupling $g_X$ of the DP.
For comparison, we report the most recent existing constraints from the collider experiments Belle II, BaBar, KLOE, CMS and LHCb, from the CCFR neutrino experiment, and indirect constraints from Fermilab measurements on the muon $g-2$.
Experimental analyses on general models, with an additional neutral vector boson connected to the SM through kinetic mixing $\epsilon$ are recast into our minimal $L_\mu - L_\tau$ framework using Eq.~\ref{eq:loop-mixing}, i.e.\ dividing $\epsilon$ by the loop factor in Eq.~\ref{eq:loop-mixing} to obtain the gauge coupling $g_X$, while rescaling the excluded cross sections by the appropriate branching ratios. 

{\it Low mass region:}
Very light DPs, with mass below the di-muon threshold, can be probed only indirectly.
In this mass region, the dominant exclusion limits are indeed given by the NA64 experiment, by the CCFR neutrino trident experiment, and by measurements of the muon $g-2$ anomaly.
The most stringent constraints from NA64 arise from a recent analysis of approximately $2\times 10^{10}$ muons generated by the CERN Super Proton Synchrotron and scattered off fixed-target atomic nuclei. 
In this analysis, a DP is radiated in a bremsstrahlung-like process from the muon after the scattering, and is searched for in events with a final state comprising a muon accompanied by missing energy which is attributed to the DP decaying into either neutrino or dark matter pairs~\cite{NA64:2024klw} (solid yellow line).
The CCFR neutrino trident experiment studies the rare process $\nu N \to \nu \mu^+\mu^-$, with the $\mu^+\mu^-$ pair produced from muon-neutrino scattering off the Coulomb field of a nucleus.
It is therefore particularly sensitive to additional contributions from neutral currents coupling to the muon flavour.
At present, measurements on SM neutrino cross sections from CCFR~\cite{Altmannshofer:2014pba} exclude gauge couplings above $g_X \simeq 10^{-3}$ for DP masses up to the di-muon threshold, and above $g_X \simeq 10^{-2}$ for DP masses of the order of 10~GeV (light grey shaded area of Fig.~\ref{fig:summary_plot}).
The purple band indicates the parameter space which would explain, with 90\% CL, the muon $g-2$ discrepancy arising at the $4.2\sigma$ level from the comparison of the Run-1 measurements at FNAL~\cite{Muong-2:2021ojo} with the SM prediction from the Muon $g-2$ Theory Initiative~\cite{Aoyama:2020ynm,Aoyama:2012wk,Aoyama:2019ryr,Czarnecki:2002nt,Gnendiger:2013pva,Davier:2017zfy,Keshavarzi:2018mgv,Colangelo:2018mtw,Hoferichter:2019mqg,Davier:2019can,Keshavarzi:2019abf,Kurz:2014wya,Melnikov:2003xd,Masjuan:2017tvw,Colangelo:2017fiz,Hoferichter:2018kwz,Gerardin:2019vio,Bijnens:2019ghy,Colangelo:2019uex,Blum:2019ugy,Colangelo:2014qya}. As mentioned in the introduction, the size and significance of the discrepancy is currently not confirmed and the band should be taken as an indication only. The muon $g-2$ constraints are mostly relevant in the low mass region, as is clear from Fig.~\ref{fig:summary_plot}. 
Constraints from collider experiments in this low mass region are only possible considering final states with missing energy or momentum.
The exclusion limit from the Belle II analysis on the 2$\mu$ plus missing energy signature with an integrated luminosity dataset of 79.7~fb$^{-1}$~\cite{Belle-II:2022yaw} is represented by the solid light pink curve.
For light DPs with masses below 1~GeV, it excludes gauge couplings above $g_X \simeq 3\cdot10^{-3}$, while for heavier DPs the sensitivity reduces exponentially to $g_X \simeq$ 1 for masses of 8~GeV.
Our optimised analysis on the 2$\mu$ plus missing energy channel (solid dark blue line), shows comparable sensitivity with respect to the current Belle II analysis, and it could test couplings above $g_X \simeq 3\cdot10^{-3}$ for DP masses up to 2~GeV, and above $g_X \simeq 10^{-1}$ for DP masses up to 8~GeV.

{\it Two muon final states:}
Analyses on final states with only two muons are, in general, the least constraining.
Searches for DPs at KLOE are sensitive to masses between 520~MeV to 990~MeV, and include both decay modes into muon and pion pairs~\cite{KLOE-2:2018kqf}.
In the context of the $L_\mu - L_\tau$ model, only the former is relevant, and the contour obtained with an integrated luminosity of 1.93~fb$^{-1}$, represented by the solid purple line, excludes gauge couplings up to $g_X \simeq 5\cdot10^{-2}$.
The dashed brown curve is derived from the CMS analysis on direct production of a light resonance decaying into a muon pair, with integrated luminosity of 96.6~fb$^{-1}$~\cite{CMS:2023hwl}.
It uses a trigger selection and muon identification optimized for the low-mass region considered, and it is sensitive to DP masses between 1.1~GeV and 7.9~GeV, excluding the region between 2.6 and 4.2~GeV, i.e.\ around the $J/\Psi$, $\Psi$(2S), and $\Upsilon$(1S) resonances.
It excludes gauge couplings as small as $g_X \simeq 2\cdot10^{-2}$ for very light DPs, with exponentially decreasing sensitivity to $g_X \simeq 1$ for heavier DPs.
The LHCb analysis on light resonances decaying into a muon pair is obtained with a dataset corresponding to an integrated luminosity of 5.5~fb$^{-1}$~\cite{LHCb:2019vmc}.
Both prompt and displaced signatures are considered, and the exclusion limits are represented by the dashed black and solid black lines respectively.
The LHCb prompt analysis shows a similar sensitivity to the analogous CMS one, but also covering the region below 1~GeV, down to the di-muon threshold, where it excludes gauge couplings $g_X \simeq 8\cdot10^{-3}$.
In the mass region between the di-muon threshold and 300~MeV, the analysis with displaced signatures gives the best sensitivity, excluding gauge couplings down to $g_X \simeq 10^{-3}$.
In comparison with these experimental results, we observe that our analysis on displaced DP decays (dashed dark blue curve) shows a shape similar to the analogous LHCb result, and a significantly improved sensitivity to gauge couplings below $g_X \simeq 10^{-4}$.

{\it Four muon final states:}
Analyses on final states with 4 muons provide strong constraints in the $L_\mu - L_\tau$ model parameter space, especially for DP heavier than 1~GeV.
The BaBar exclusion curve (solid orange line) has been obtained from the analysis of the 4$\mu$ prompt signature, considering the dataset corresponding to an integrated luminosity of 514~fb$^{-1}$, taken at the $\Upsilon$(4S) resonance and in the neighborhood of the $\Upsilon$(3S) and $\Upsilon$(2S) peaks~\cite{BaBar:2016sci}.
It covers the mass region above the di-muon threshold, excluding gauge couplings below $g_X \simeq 10^{-3}$ up to masses of 2~GeV, while for heavier DPs the sensitivity reduces exponentially to $g_X \simeq 1$ for masses of 8~GeV.
An analogous analysis from Belle II with a dataset corresponding to an integrated luminosity of 643~fb$^{-1}$~\cite{Belle:2021feg} shows a similar sensitivity (solid red line).
Belle II performs slightly worse than BaBar for DP lighter than 2~GeV, while marginally improving BaBar exclusions for DPs heavier than 2~GeV.
Very recently, an analysis of 178~fb$^{-1}$ in the four muon final state has been released~\cite{Belle-II:2024wtd} by Belle II. The results (solid dark green line) show a sensitivity similar to BaBar and previous Belle II analyses, and further improving upon CCFR. The combination of all three results sets stringent exclusion limits for DP masses above 3~GeV.
The CMS analysis of the four muons prompt final state with an integrated luminosity of 77.3~fb$^{-1}$~\cite{CMS:2018yxg} covers only DPs heavier than 5~GeV (solid brown line), and provides the strongest exclusion down to $g_X \simeq 5\cdot10^{-3}$.
Our results on the four muons prompt signature (dot-dashed dark blue line) show a significant improvement with respect to BaBar and previous Belle II analyses, increasing the gauge coupling exclusion by roughly a factor 2.
For DPs as heavy as 5-6~GeV, our analysis is sensitive to gauge couplings down to $g_X \simeq 5\cdot10^{-3}$, and comparable with the CMS high mass analysis.

A special mention is deserved by our study of the monophoton channel, whose sensitivity curve is given by the dotted dark blue line.
The interpretation of this signature within the $L_\mu - L_\tau$ framework shows a remarkable potential for constraining specifically the minimal construction we have considered.
The underlying process is indeed very sensitive to the kinetic mixing $\epsilon$, which in our case is not a free parameter, but instead directly connected to the gauge coupling $g_X$ through the kinetic mixing loop factor, $f(m_\mu,m_\tau,q)$.
This analysis predicts the strongest sensitivity for DP masses below the di-muon threshold, excluding down to $g_X \simeq 4\cdot10^{-4}$, which is comparable or better than the strongest indirect bound from the muon $g-2$.
For heavier DPs, the sensitivity of the monophoton analysis closely follows the one of the four muons prompt analysis.

\section{Conclusions}
\label{sec:conclusions}
In this work we have explored the sensitivity of the Belle II experiment to a minimal dark photon (DP) model with an underlying $L_\mu - L_\tau$ extra U(1) gauge group.
Assuming the 50~ab$^{-1}$ target luminosity of Belle II, we projected the sensitivity contours of multiple signatures on the model's parameter space which, in its minimal configuration, reduces to the DP's mass-coupling plane $m_X - g_X$.
We considered the most promising signatures providing the best sensitivity at lepton colliders.
They include final states where the DP appears as missing energy, such as the monophoton and two muons plus missing energy channels, and also multi-lepton final states where a pair of muons originates from the DP decay.
The latter can arise as a prompt signature, i.e.\ as a resonant excess in a narrow invariant mass bin, or as a displaced signature, where the long-lived DP decays at a measurable distance with respect to the principal interaction point.
For the analysis of this particular signature, we included the simulation of displaced decay events in both signal and background samples.
In all the simulated events we considered the detector's fiducial acceptance and reconstruction efficiencies for each particle, and for the VXD a resolution of 5~$\mu$m is assumed.
The sensitivity projections are compared to the most recent results from direct searches of DPs at lepton colliders from the BaBar, Belle and KLOE experiments, at the LHC from the CMS and LHCb experiments, and to indirect probes using CCFR neutrino trident production data and the muon anomalous magnetic moment $g-2$.

We find that with its ultimate 50~ab$^{-1}$ target dataset, the Belle II sensitivity on light DPs will improve on all existing exclusion limits.
For very light DPs, i.e.\ below the two muon decay threshold, we observe the strongest sensitivity in the monophoton channel, which will overlap or extend the sensitivity band of the muon $g-2$ indirect probe.
When the DP is heavy enough to decay into a pair of muons, multi-leptonic final states will provide the most sensitive signatures, and only for heavy DPs ($m_X>8$~GeV) the LHC sensitivity becomes dominant.

The displaced decay signature in particular reveals a significant potential for the searches of narrow, long-lived DPs.
Despite its limited sensitivity region, it allows to probe parameter regions not accessible through other signatures.
The shape of the sensitivity contour is strongly dependent on the choice of the minimal VXD spacial resolution.
In the minimal $L_\mu - L_\tau$ model, the typical DP decay displacements are indeed comparable with the VXD resolution.
Assuming 5~$\mu$m, we showed that the displaced analysis can test DP couplings $g_X \sim 10^{-4}$ and below, improving current bounds by one order of magnitude.
While this assumption may seem optimistic, future developments in the muon track reconstruction may allow this level of precision or even better.
In fact, even small further improvements of the VXD resolution would lead to critical reductions of the SM fake displaced background rates.
This in turn would significantly increase the sensitivity of long-lived particles searches and eventually turn this signature into a very promising channel for the discovery of such BSM signals.

While $L_\mu - L_\tau$ DPs would also decay into tau pairs with a significant rate, we did not focus on signatures involving tau leptons, because the experimental challenges connected to their reconstruction generally decrease the overall sensitivity in comparison with purely muonic final states.
Yet, we stress that a combined fit of different search channels can help to identify the specific characteristics of the new physics model, such as an additional BSM symmetry group.
For instance, the equivalence of DP decay rates into muons and taus would provide strong indications of an underlying $L_\mu - L_\tau$ symmetry.
Similarly, assessing the BSM contribution to the monophoton channel gives a direct handle on the kinetic mixing parameter $\epsilon$.
This parameter is very sensitive to the presence of additional hidden states, and can help to distinguish between BSM models with extra gauged $U(1)_X$ symmetries and generic models featuring additional vector boson resonances.

The analysis methods presented in this work are very general and can be applied to various BSM scenarios across different collider experiments.
The generic procedure, to determine for each signature the optimal kinematic cuts maximising the experimental sensitivity to a new physics signal, can be applied to different detector configurations and BSM models.
In addition, our analysis of displaced signatures, which includes the simulation of both signal and background contributions, can be easily adapted to specific experimental environments and BSM constructions predicting long-lived particles.
To fully assess the ultimate experimental sensitivity, more accurate theoretical predictions including higher order corrections and extra radiation are required, together with detailed detector simulations.
While this level of analysis is beyond the scope of our current study, we plan to address some of these aspects in future work.

\section*{Acknowledgements}
We thank Prof. Gianluca Inguglia for useful discussions on the Belle II detector. CB is supported by an STFC PGR studentship. JF and TT were supported by the STFC Consolidated Grant ST/T000988/1 and TT currently by ST/X000699/1.
JF acknowledges financial support from ICSC~– Centro Nazionale di Ricerca in High Performance Computing, Big Data and Quantum Computing, funded by the European Union – NextGenerationEU.

\bibliography{bibliography}

\end{document}